\documentclass[twocolumn]{openjournal}
\usepackage{natbib}
\usepackage{graphicx}
\usepackage{graphicx,amsmath,amssymb,amstext}
\usepackage{amsbsy,amsfonts,amsthm,color}
\usepackage{caption}
\usepackage{subcaption}
\usepackage{xcolor}
\usepackage{graphicx}
\usepackage{lipsum}
\usepackage{bm}
\usepackage{cancel}

\usepackage{tikz}
\usetikzlibrary{shapes,arrows,shadows,fit}
\usetikzlibrary{positioning}
\usetikzlibrary{bayesnet}
\definecolor{ourgreen}{HTML}{06D6A0}
\definecolor{ourblue}{HTML}{118AB2}

\usepackage{hyperref}
\hypersetup{colorlinks,linkcolor=ourblue,citecolor=ourblue,urlcolor=ourblue}

\newcommand{\almanac}{\textsc{Almanac}}

\newcommand{\citepCore}{(Sellentin, Loureiro et al., \textit{in prep})}
\newcommand{\citeCore}{Sellentin, Loureiro et al., \textit{in prep}}

\newcommand{\imagunit}{\ensuremath{\textrm{i}\mkern1mu}}

\newcommand{\half}{\frac{1}{2}}

\newcommand{\transpose}{\ensuremath{{\textrm{T}}}}

\newcommand{\sfL}{{\ensuremath{\sf{L}}}}
\newcommand{\sfN}{{\ensuremath{\sf{N}}}}

\newcommand{\sfC}{{\ensuremath{\sf{C}}}}

\newcommand{\sfA}{\ensuremath{{\sf{A}}}}

\newcommand{\sfJ}{\ensuremath{{\sf{J}}}}

\newcommand{\mathd}{\ensuremath{\mathrm{d}}}

\newcommand{\fata}{\ensuremath{\boldsymbol{a}}}
\newcommand{\fatd}{\ensuremath{\boldsymbol{d}}}

\newcommand{\calP}{\ensuremath{\mathcal{P}}}

\newcommand{\calL}{\ensuremath{\mathcal{L}}}

\newcommand{\calG}{\ensuremath{\mathcal{G}}}

\newcommand{\calN}{\ensuremath{\mathcal{N}}}
\newcommand{\fatx}{\ensuremath{\boldsymbol{x}}}

\newcommand{\sfK}{\ensuremath{{\sf{K}}}}

\newcommand{\sfY}{\ensuremath{{\sf{Y}}}}

\newcommand{\sfG}{\ensuremath{{\sf{G}}}}

\newcommand{\pos}{\ensuremath{{\hat{n}}}}

\newcommand{\newtpot}{\ensuremath{{\Psi}}}
\newcommand{\lpot}{\ensuremath{{\Tilde{\Psi}}}}
\newcommand{\nlp}{\ensuremath{{\psi}}}

\definecolor{ork}{rgb}{0.9,0.1,0.3}
\definecolor{grbl}{rgb}{0.3,0.6,0.7}
\definecolor{bleu}{rgb}{0,0.5,0.5}

\usepackage{soul}

\usepackage{xcolor}

\begin{document}

\journalinfo{The Open Journal of Astrophysics}
\shorttitle{Almanac: Weak Lensing Power Spectra and Map Inference}

\title{Almanac: Weak Lensing power spectra and map inference on the masked sphere}

\author{Arthur Loureiro$^{1,2,3}$}
\author{Lorne Whiteway$^{2}$}
\author{Elena Sellentin$^{4,5}$}
\author{Javier Silva Lafaurie$^{4,5}$}
\author{Andrew H. Jaffe$^{3}$}
\author{Alan F. Heavens$^{3}$}

\affiliation{$^1$ Institute for Astronomy, University of Edinburgh, Royal Observatory, Blackford Hill, Edinburgh EH9 3HJ, UK}
\affiliation{$^2$ Department of Physics and Astronomy, University College London, Gower Street, London WC1E 6BT, UK}
\affiliation{$^3$ Astrophysics Group and Imperial Centre for Inference and Cosmology (ICIC), Blackett Laboratory, Imperial College London, Prince Consort Road, London SW7 2AZ, UK}
\affiliation{$^4$ Mathematical Institute, Leiden University, Snellius Gebouw, Niels Bohrweg 1, NL-2333 CA Leiden, The Netherlands}
\affiliation{$^5$ Leiden Observatory, Leiden University, Oort Gebouw, Niels Bohrweg 2, NL-2333 CA Leiden, The Netherlands}

\email{arthur.loureiro@ed.ac.uk}
\email{lorne.whiteway@star.ucl.ac.uk}
\email{sellentin@strw.leidenuniv.nl}
\email{silvalafaurie@strw.leidenuniv.nl}
\email{a.jaffe@imperial.ac.uk}
\email{a.heavens@imperial.ac.uk}

\begin{abstract}

We present a field-based signal extraction of weak lensing from noisy observations on the curved and masked sky. We test the analysis on a simulated Euclid-like survey, using a Euclid-like mask and noise level. To make optimal use of the information available in such a galaxy survey, we present a Bayesian method for inferring the angular power spectra of the weak lensing fields, together with an inference of the noise-cleaned tomographic weak lensing shear and  convergence (projected mass) maps. The latter can be used for field-level inference with the aim of extracting cosmological parameter information including non-gaussianity of cosmic fields. We jointly infer all-sky $E$-mode and $B$-mode tomographic auto- and cross-power spectra from the masked sky, and potentially parity-violating $EB$-mode power spectra, up to a maximum multipole of $\ell_{\rm max}=2048$. We use Hamiltonian Monte Carlo sampling, inferring simultaneously the power spectra and denoised maps with a total of $\sim 16.8$ million free parameters. The main output and natural outcome is the set of samples of the posterior, which does not suffer from leakage of power from $E$ to $B$ unless reduced to point estimates. However, such point estimates of the power spectra, the mean and most likely maps, and their variances and covariances, can be computed if desired. 

\end{abstract}

\maketitle

\section{Introduction}\label{Sec:Intro}

Cosmic shear analysis has developed over the past few decades into a powerful cosmological probe for studying the large-scale structure of the Universe \citep{BartelSchneid,2008-Munshi-Review,MartinReview}. Photons from distant galaxies travel towards us along geodesic paths that are distorted by the gravitational potential of dark and baryonic matter along the way. The result is a coherent distortion in the observed shapes of background galaxies: a weak gravitational lensing effect that shears and magnifies objects \citep{BartelSchneid}. The shape distortions are sensitive to the total matter content of the Universe and its clustering \citep{2021-Hall-S8}, and so this effect makes cosmic shear an effective probe for studying the growth of structure and the geometric evolution of the Universe \citep{2009-Heavens-Review}. Measuring how the gravitational potential along the line-of-sight affects the ellipticities of galaxies in photometric galaxy surveys also allows us to map the invisible distribution of dark matter around galaxies \citep{1993-Kaiser-MassMapping,2002-Marshall-MassMap,2016-Lanusse-MassMap,2021-Jeffrey-DES-MASSMAP,2022-Wallis} and to understand the structure of the cosmic web \citep{2012-Rossi-CosmicWeb,2015-Joachimi,2018-Codis}.

Recent years have seen an impressive increase in both the quality and quantity of weak lensing data and analysis techniques. Current photometric surveys such as the Kilo Degree Survey (KiDS; \citealt{2021-Heymans-KiDS1000-Cosmology, 2021-Asgari-2ptsK1000,2021-Loureiro-Kids-PCL}), the Dark Energy Survey (DES; \citealt{2021-DES-Y3-3x2pt, 2021-Amon-DES-Y3-Shear, 2021-Secco-DES-Y3-Shear}), and the Subaru Hyper Supreme-Cam survey (HSC; \citealt{HSC, 2019-Hikage-HSC}) have, together, revolutionised cosmology by using cosmic shear to challenge its standard model. Contemporary weak lensing experiments infer a growth of structure parameter ($S_8$) that is in tension with the value inferred from the cosmic microwave background (CMB) \citep{2013-Heymans,2014-MacCrann-tensions,2020-Lemos-tension,2022-AmonEfstat}. 

New data from forthcoming cosmic shear surveys such as Euclid \citep{Euclid} and LSST \citep{LSST}, combined with improved data analysis techniques, could shed light on this tension and help us better understand the geometry and growth of structure in the Universe. These future weak lensing experiments will measure the shapes of billions of galaxies across large portions of the sky with unprecedented accuracy and precision. The question `how to extract optimally the cosmic shear information from these data?' is hotly debated in the field \citep{2014-Kilbinger-Athena,2015-Jarvis-TreeCore,2019-Alonso-Namaster,2021-Nicola,2021-Asgari-2ptsK1000, 2021-Porqueres-WL-1,2022-Porqueres-2}.

The vastly increased precision expected from Euclid and LSST raises the question of whether the traditional approach of summary statistics such as correlation functions and pseudo-power spectrum estimates are accurate enough for robust inference. There is some evidence that shear correlation functions fail \citep{2009-Hartlap,2018-Sellentin,SHH18}, while pseudo-C$_\ell$ estimates are close to gaussian \citep{2022-Upham-PClCov}. Avoiding having to assume the gaussianity of such summaries is a major benefit of field-level inference approaches, whose goal is to sample correctly from the likelihood of the (close to) raw field-level data  

There are two general approaches to field-level inference of cosmic shear: one firmly based in a cosmological model with a physical gravity model, the other (the approach taken in this paper) more agnostic, inferring the underlying shear maps and their two-point statistics in a cosmology-independent way with minimal assumptions.

In an example of the former, model-based approach, \cite{2022-Porqueres-2} showed that field-level inference can strongly lift the standard `banana' cosmic shear parameter degeneracy (the degeneracy between the amplitude of matter density fluctuations $\sigma_8$ and the total matter density $\Omega_{\rm m}$). By extending the \textsc{Borg} framework \citep{2013-Jasche-BORG1,2019-Jasche-BORG2}, which includes a full gravity model, sampling of the initial gaussian density field is performed and forward modelled to the tomographic shear fields, where the likelihood is applied. This is a very powerful framework, but it ties the inference to a particular cosmological model.
 
 The approach of this work --- as embodied in our algorithm, called \almanac{} --- is complementary, inferring the shear maps and the power spectra of the (denoised, unmasked) fields in a cosmology-independent way. The samples of maps and power spectra produced by \almanac{} are not tied to a particular model (beyond a few basic symmetry assumptions), and can subsequently be used for parameter inference (although the goal of this paper is simply to draw map and power spectra samples). The main challenge is efficient scaling with the highest multipole probed, so we concentrate on a relatively simple 2-bin tomographic example. More details of \almanac{} are presented in a companion paper \citepCore{}. 
 
 The paper is structured as follows: weak lensing theory in section \ref{Sec:WLtheory} is followed by the hierarchical data model in section \ref{Sec:Methods}, and the Cholesky reparameterisation in section \ref{sec:cholesky}. Simulations are described in section \ref{Sec:Sims}, and results and convergence tests presented in section \ref{Sec:Results}. We give our conclusions in section \ref{Sec:Discussion}.

\section{Weak lensing theory}
\label{Sec:WLtheory}

Galaxies are biased tracers of the dark matter density field. By contrast, weak distortions in the observed shapes of background galaxies (caused by the deflection of light by foreground structures) are an excellent tracer of the projected matter field. This section outlines the formalism behind using this weak lensing as a cosmological probe. Further details can be found in reviews by \cite{BartelSchneid} and \cite{MartinReview}.

The usual approximation of the lens equation maps the unperturbed two-dimensional source angular position $\mathbf{\theta}_s$ to the observed position $\mathbf{\theta}_{\text{obs}}$ via a mapping $\sfA_{ij}$: $\theta_{s,i} \approx \sfA_{ij}\theta_{\text{obs},j}$. This mapping can be expressed in terms of the effective lensing potential $\lpot{}$ as
\begin{equation}
    \sfA_{ij} = \delta_{ij} - \partial_i\partial_j\lpot{},
    \label{Eq:JacPotential}
\end{equation}
where $\lpot{}$ is a weighted projection along the line-of-sight of the three-dimensional Newtonian potential $\newtpot{}$ \citep{1992-KaiserLensing}:
\begin{equation}
    \lpot{}(\chi_{\rm s},\pos{}) = 2\int_0^{\chi_s} d\chi\frac{f_K(\chi_{\rm s} - \chi)}{f_K(\chi)f_K(\chi_s)}\newtpot{}(\chi,\pos{})\, .
\label{Eq:lensing_potential}
\end{equation}
Here $\chi_{\rm s}$ is the comoving distance to the lensing source, and $f_K$ is related to the curvature $K$ of the universe (for a flat universe $f_{K=0}(\chi) = \chi$, where $\chi$ is a comoving distance variable).  We also define the tomographic-bin averaged quantity
\begin{equation}
    \lpot{}(\pos) = \int d\chi \ \bar{n}(\chi) \lpot{}(\chi, \pos)
\end{equation}
where $\bar{n}(\chi)$ is the normalised density of sources (averaged over the sky); $\lpot{}$ is a scalar (i.e., spin-0) field on the sky.

The mapping in Eq.~\eqref{Eq:JacPotential} can also be written in terms of the spin-0 lensing convergence field $\kappa$ and the complex spin-2 shear field $\bm\gamma = \gamma_1 + \imagunit{}\gamma_2$:
\begin{align}
     \mathsf{A} = \left(
\begin{array}{cc}
1 - \kappa - \gamma_1 & -\gamma_2 \\
-\gamma_2 & 1 - \kappa + \gamma_1
\end{array}\right).
\end{align}

The potentially observable quantities $\kappa$ and $\gamma$ are related to the lensing potential $\lpot{}$ by
\begin{equation}
    \kappa  = \frac{1}{4}(\eth\bar{\eth} + \bar{\eth}\eth)\lpot{}
    \label{Eq:kappa_realspace}
\end{equation}
and
\begin{equation}
    \boldsymbol{\gamma} = \gamma_1 + \imagunit{}\gamma_2 = \half\eth\eth\lpot{},
    \label{Eq:gamma_realspace}
\end{equation}
where $\eth$ and $\bar{\eth}$ are the spin-raising and spin-lowering differential operators (see, e.g., \cite{2005-Castro} for details).

An arbitrary spin-$s$ function $f$ on the sphere may be represented in the basis of spin-$s$ spherical harmonic functions ${}_s Y_{\ell m}$:
\begin{equation}
    f(\pos) = \sum_{\ell m} f_{\ell m} \ {}_s Y_{\ell m}(\pos)
\end{equation}
where
\begin{equation}
    f_{\ell m} = \int d\Omega \ f(\pos) \ {}_s Y_{\ell m}^{*}(\pos).
\end{equation}
In this basis Eqs.~\eqref{Eq:kappa_realspace} and \eqref{Eq:gamma_realspace} become
\begin{align}
    \kappa_{\ell m} & = - \half\ell(\ell+1)\, \lpot{}_{\ell m}\quad \textrm{and} \label{Eq:Kappa_lm}\\
     \gamma_{\ell m} & = \half\sqrt{(\ell -1)\ell(\ell+1)(\ell+2)}\, \lpot{}_{\ell m}\,, \label{Eq:gamma_lm}
\end{align}
respectively, and hence
\begin{equation}
     \gamma_{\ell m}  = -\sqrt{\frac{(\ell-1)(\ell+2)}{\ell(\ell+1)}} \kappa_{\ell m}.
     \label{Eq:gamma_lm_from_kappa}
\end{equation}

The shear field may also be decomposed into $E$- and $B$-modes, with coefficients 
\begin{equation}
    \label{Eq:Elm}
    E_{\ell m} = - \half \int {\rm d}\Omega  \left[\gamma(\pos{})\,_{+2}Y^*_{\ell m}(\pos{}) + \gamma^*(\pos{})\,_{-2}Y^*_{\ell m}(\pos{})\right]
\end{equation}
and
\begin{equation}
    \label{Eq:Blm}
    B_{\ell m} = \frac{\imagunit{}}{2} \int {\rm d}\Omega \left[ \gamma(\pos{})\,_{+2}Y^*_{\ell m}(\pos{}) - \gamma^*(\pos{})\,_{-2}Y^*_{\ell m}(\pos{})\right]\, .
\end{equation}
The integrals are taken over the full sphere. The advantage of this decomposition is that to a good approximation, cosmic shear produces only $E$-modes \citep[e.g.][]{2005-Schneider,2017-Bartelmann}, and the presence of $B$-modes can indicate systematic errors in the data \citep[e.g.][]{2006-Heymans,2019-Asgari-BModes}. These definitions use the $H=1$ sign convention (in the notation of \cite{2005PhRvD..71h3008L}), as is appropriate for \textsc{HEALPix}; one consequence is that $\gamma=-(E + \imagunit{} B)$.

Galaxy surveys directly observe neither shear nor convergence, observing instead galaxy shape ellipticities $\bm\epsilon$ (and optionally sizes and/or fluxes). These shapes are noisy estimates of the shear field in the weak lensing regime ($\kappa \ll 1$). In this regime one has, for a large number of galaxies, $\langle \bm\epsilon \rangle \approx \bm\gamma/(1-\kappa) \approx \bm\gamma$. Both shape noise (due to the random intrinsic shapes of galaxies) and measurement noise (a smaller effect than shape noise) are taken into account in the standard deviation $\sigma_{\epsilon}$ of the distribution of the ellipticities; this distribution is usually assumed to be approximately gaussian for a sufficiently large area density of sources \citep{2010-JoachimiBridle, 2013-VanWearbeke, 2016-Alsing,2018-Chang-MassMapDES}. Intrinsic alignments can also introduce correlations of ellipticities \citep[e.g.,][]{2000-HRH,2001-Catelan,2015-Kirk-IntrAlignments, 2015-Kiessling,2015-Joachimi,2015-Blazek-IA}, and this effect will be included in the inferred power spectra.

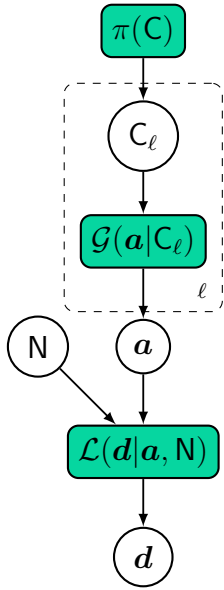
\begin{figure}
    \begin{center}
    \large{\begin{tikzpicture}[node distance = 1.4cm, auto]

	    \pgfdeclarelayer{background}
	    \pgfdeclarelayer{foreground}
	    \pgfsetlayers{background,main,foreground}
	    \tikzstyle{prob}=[draw, thick, text centered, rounded corners, minimum height=1em, minimum width=1em, fill=ourgreen]
	    \tikzstyle{var}=[draw, thick, text centered, circle, minimum height=1em, minimum width=1em, fill=white]

        \node[prob](prior){$\pi(\sfC)$};
        \node[var](Cls)[below of= prior]{$\sfC_\ell$};
        \node[prob](Palms)[below of= Cls]{$\calG(\fata | \sfC_\ell)$};
        \plate[inner sep=.2cm, yshift=.02cm, dashed] {plate1} {(Cls) (Palms)} {$\ell$};
        \node[var](alms)[below of= Palms]{$\fata$};
        \node[prob](like)[below of= alms]{$\calL(\fatd | \fata, \sfN)$};
        \node[var](noise)[left of= alms]{$\sfN$};
        \node[var](data)[below of= like]{$\fatd$};
        
        \path [draw, line width=0.7pt, arrows={-latex}] (prior) -- (Cls);
        \path [draw, line width=0.7pt, arrows={-latex}] (Cls) -- (Palms);
        \path [draw, line width=0.7pt, arrows={-latex}] (Palms) -- (alms);
        \path [draw, line width=0.7pt, arrows={-latex}] (alms) -- (like);
        \path [draw, line width=0.7pt, arrows={-latex}] (noise) -- (like);
        \path [draw, line width=0.7pt, arrows={-latex}] (like) -- (data);

    \end{tikzpicture}}
    \end{center}
    \caption{
    Directed acyclic graph representing the Bayesian hierarchical model in Eq.~\eqref{Eq:Cond_Posterior} for maps and angular power spectra inference. Begin by drawing the angular power spectra $\sfC_{\ell}$ from a prior distribution $\Pi(\sfC_{\ell})$. Then draw a set of spherical harmonics coefficients for the shear fields $\fata$, conditional upon the angular power spectra. Finally draw a set of noisy tomographic cosmic shear fields given the signal, $\fata$, and the noise covariance, $\sfN$, based on the realisations from the previous steps.}
    \label{Fig:DAG}
\end{figure}

\section{The posterior and HMC sampling}
\label{Sec:Methods}
\subsection{Joint posterior of fields and their covariance}
The posterior from which we draw samples consists of the full sky maps in harmonic space $\fata$ and their power spectra $\sfC$, given the data $\fatd$ and the noise covariance $\sfN$. Fig.~\ref{Fig:DAG} shows a directed acyclic graph for the model presented in this section. Here $\fata$ and $\fatd$ are concatenations of the spherical harmonic coefficients and the field values from different tomographic bins into a single map parameter vector and a single data vector. The power spectrum $\sfC$ contains auto- and cross-power spectra across tomographic bins, with $E$- and $B$-mode correlations, as well as the parity-violating $EB$-mode cross-correlations. For more details, see the companion paper \citepCore{}. The posterior is
\begin{equation}
    \calP(\sfC, \fata | \fatd, \sfN) \propto \calL(\fatd | \fata, \sfN)\calG(\fata | \sfC)\pi(\sfC)\, .
    \label{Eq:Cond_Posterior}
\end{equation}
The first factor is the probability at field level:
\begin{equation}
    \calL(\fatd | \fata, \sfN) \propto \exp \left(-\half (\fatd - \sfY\fata)^{\rm T}\sfN^{-1}(\fatd - \sfY\fata) \right) \, .
    \label{Eq:Likelihood}
\end{equation}
The second factor assumes a gaussian prior for the spherical harmonic coefficients:
\begin{equation}
    \calG(\fata|\sfC) = \frac{1}{\sqrt{|2\pi \sfC|}} \exp\left( -\half \fata^{\transpose{}} \sfC^{-1}\fata \right).
\end{equation}
This factor merits some discussion. The aim of the method is to infer the power spectra and the maps in a cosmology-independent way, making no assumptions about the gravity model. Thus, in contrast to BORG-WL \citep{2021-Porqueres-WL-1,2022-Porqueres-2}, we do not sample the initial gaussian field and then forward model. Instead, we want the least informative prior for the coefficients, for a given power spectrum, with no reliance on a model-dependent knowledge (in principle) of the statistics of the shear fields. This least informative, maximum-entropy prior is a gaussian \citep{jaynes03}, which ensures that we are not using any model-dependent information.
This does not imply that the fields themselves are realisations of a zero-mean gaussian with isotropic covariance, nor that a map created from the spherical harmonic coefficients of a given posterior sample is a realisation of such a process, since these are conditioned on the data. This is similar to the known behaviour of Wiener filters: where the data are highly constraining, the posterior will have small variance around the data, but where they are noisy, the variance will be dominated by the prior information about the signal.

The last factor in the posterior is a prior on the angular power spectra. In all cases we set the prior to zero outside the set of positive definite matrices, enforced by the coordinate choices described in Section~\ref{sec:cholesky}. We then assign a constant prior, $\pi(\sfC) = \text{const}$, but note that this choice is essentially cosmetic. Any use of these results for further analysis will require a prior on cosmological parameters $\theta$, corresponding to a delta-function distribution $\pi(\theta, \sfC) = \pi(\theta)\delta(\sfC-\sfC[\theta])$. In this case the constant prior is equivalent to sampling from the likelihood function itself, which can then be modulated by such a cosmological prior.

More generally, we could choose the prior to be a power of the determinant, $\pi(\sfC) = |\sfC|^q$, enforcing different information about the scaling of the spectra. A Jeffreys prior corresponds to $q = -(N_p+1)/2$ (where $\sfC_{\ell}$ has $N_p\times N_p$ entries), as discussed in \cite{SH}; a frequency-matching choice of $q$ can also be made \citep{2022-Percival} if desired. All of these would lead to different detailed inferences, such as regarding the significance of non-zero power of a particular $\sfC_\ell$.

\subsection{Hamiltonian Monte Carlo}
Our aim is to infer the denoised maps and power spectra from simulated input data, using the Bayesian hierarchical model depicted in Fig.~\ref{Fig:DAG}. Only a few samplers can handle such a high-dimensional parameter space. Gibbs sampling \citep{Geman1984,Casella1992} as used by \cite{Wandelt2004, 2004-Eriksen, 2007-Larson,2016-Alsing,2017-Alsing,2022-Colombo} can generate samples using the conditional distributions $\calP(\fata|\sfC,\fatd)$ and $\calP(\sfC|\fata,\fatd)$; however, this method is probably too inefficient to sample the full posterior distribution in Eq.~\eqref{Eq:Cond_Posterior} for upcoming weak lensing analyses. 

Instead, we take an approach similar to that used by most field-level inference works \citep{2010-Jasche,2013-Jasche-BORG1,2015-Leclercq,2019-Jasche-BORG2, 2021-Porqueres-WL-1,2022-Porqueres-2}: we implement a tuned version of the HMC algorithm. This section outlines the key details; see the \almanac{} core paper \citepCore{} for more information.

In HMC \citep{2001-Hanson-HMC,2007-Hajian-HMC,2011-Neal} we view the negative logarithm of the posterior density as a potential energy
\begin{equation}
    \nlp{}(\fata, \sfC|\fatd) = - \ln \calP(\fata, \sfC |\fatd ).
    \label{Eq:Potential_Generic1}
\end{equation}
We then augment the parameter space with momentum variables drawn from a gaussian distribution, whose negative logarithm can be viewed as kinetic energy. A Hamiltonian trajectory of the resulting dynamical system is then computed using a leapfrog integrator. At the end of a trajectory, a Metropolis-Hastings step \citep{1953-Metropolis, 1970-Hastings} decides if the trajectory is accepted (this corrects for any non-conservation of energy by the leapfrog integrator). If it is accepted, the new sample is the trajectory's end point; if it is rejected, the new sample is the trajectory's start point. In either case, the momenta are resampled and another trajectory is computed (starting at the new sample). By iteration, the sampler builds up a Monte Carlo Markov Chain.

Step sizes and other parameters of the leapfrog integrator must be carefully tuned; see the companion paper \citepCore{} for details of this tuning.
The sampling is challenging, since the shape of the posterior in the $\fata \times \sfC$ plane exhibits a `stingray' geometry \citep{2000-Neal,2013-Betancourt} that results in large correlation lengths and slow convergence; a reparametrisation of the parameter space, as discussed in the following section, can improve convergence. 
\section{Cholesky decomposition coordinates}
\label{sec:cholesky}
\begin{figure}%
    \centering
    \subfloat[\centering Matrix-log parameters, $\{\fata, \sf{G} \}$]{{\includegraphics[width=.50\textwidth]{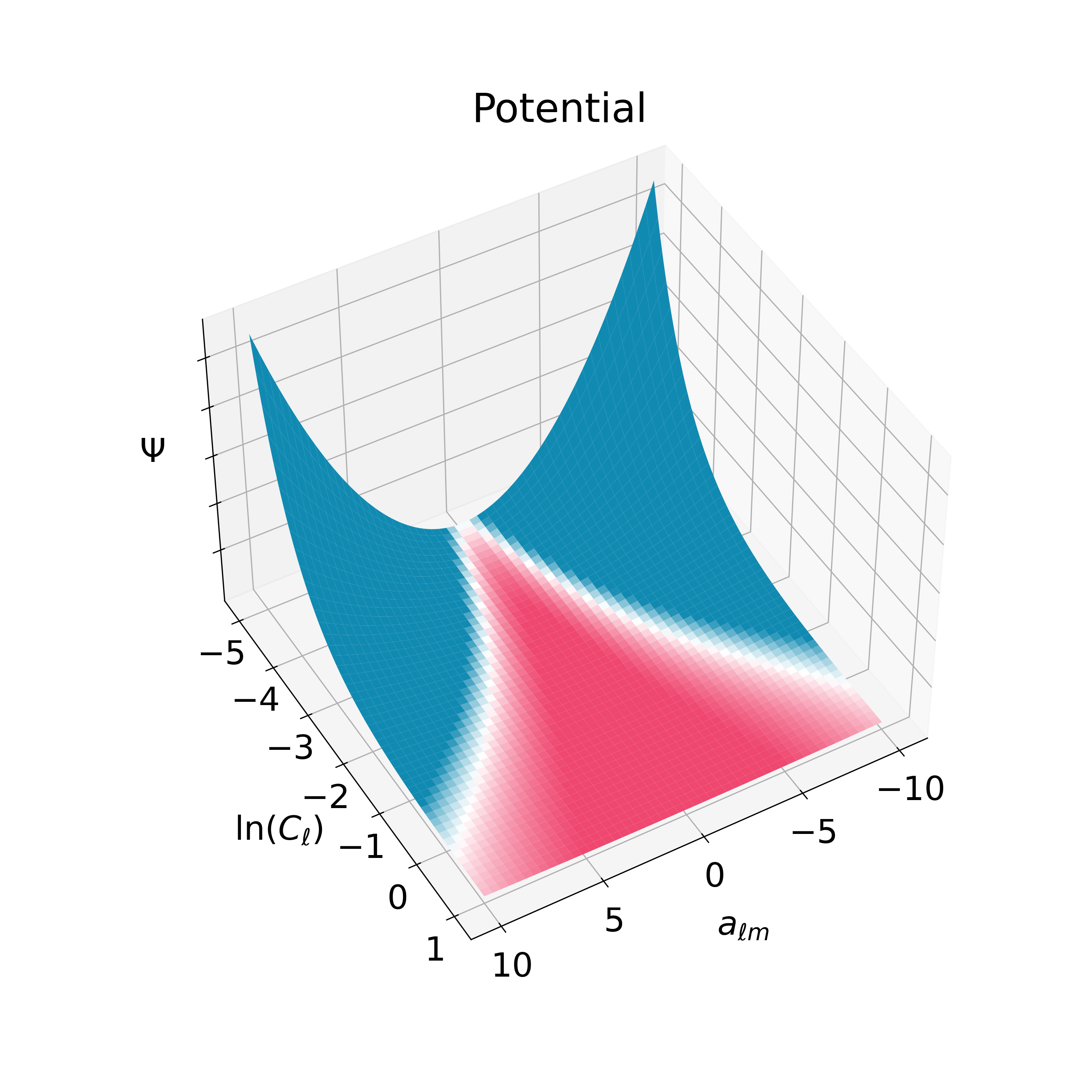}}\label{sting}}%
    \qquad
    \subfloat[\centering Cholesky decomposition parameters, $\{\fatx, \sfK \}$]{{\includegraphics[width=.50\textwidth]{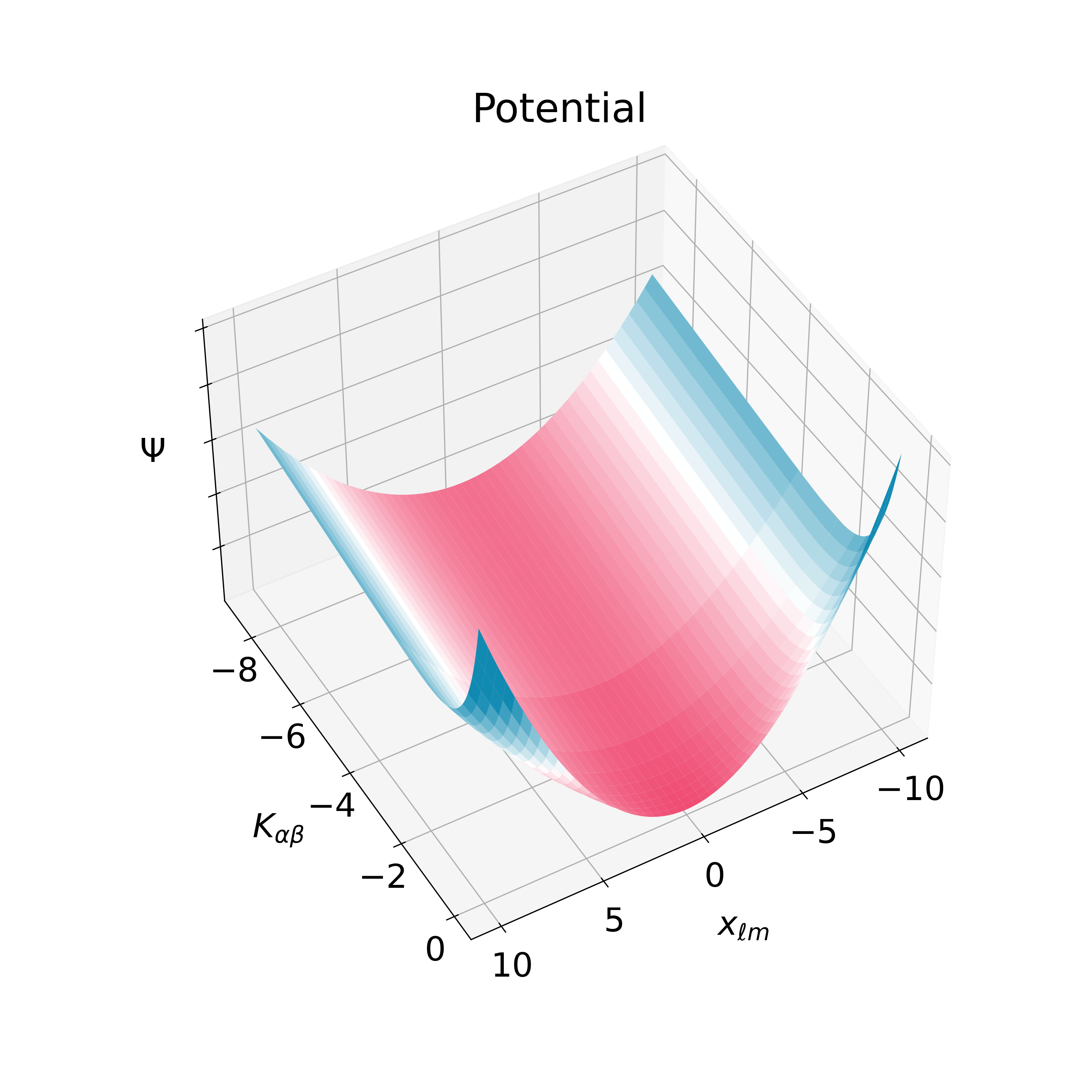} }}%
    \caption{A toy example illustrating the geometry of the potential density in \textit{(a)} the matrix-log parameterisation  $\{\fata, \sf{G} \}$ and in \textit{(b)} the Cholesky decomposition parameterisation $\{\fatx, \sfK \}$. This toy example shows a slice in the map $\times$ angular power spectra dimensions for the negative log-posterior (potential) illustrated by the DAG in Fig.~\ref{Fig:DAG} using different coordinate systems. The potential density varies from low (pink) to high (blue) and values have been omitted since the normalisation is arbitrary. The matrix-log parameterisation has a stingray shape, also known as Neal's Funnel \citep{2000-Neal}; this posterior shape is known to be difficult to sample from.}%
    \label{stingray}%
\end{figure}

The particular challenge of tomographic weak lensing is that the covariance matrix, although block-diagonal in $\ell$ and $m$, is dense for the $E$ and $B$ correlations across $N_{\rm b}$ tomographic bins. Thus there are dense sub-block matrices of size $2N_{\rm b} \times 2N_{\rm b}$, with strong correlations between neighbouring bins. We cannot straightforwardly sample the elements of $\sfC$ since the HMC random walk can easily wander out of the subset of positive-definite matrices. To handle this we instead explored sampling either a) the matrix logarithm $\sfG$ (where $\sfC= \exp(\sfG)$) 
or b) the `diagonal-log' $\sfK$ (see Eq.~\eqref{Eq:DiagonalLog_L}) of the Cholesky factor $\sfL$ (where $\sfC=\sfL\sfL^{\rm T}$ with $\sfL$ lower-triangular) of the covariance matrix. Figure ~\ref{stingray} shows, for a toy example, the geometry of the potential (the negative of the logarithm of the posterior density) using both parameterisations. For the strong correlations inherent in cosmic shear, the Cholesky decomposition was found to lead to chains with shorter correlation lengths (indeed, the matrix log formalism failed badly for this application). See below for further details of the Cholesky reparametrisation here.

The exact shape of the posterior density of course depends on the coordinates used. When parameterising using natural coordinates $\{\fata, \sfC\}$ the posterior density has an irksome shape in the vicinity of $\fata = 0, \sfC = 0$. At such points there is poor agreement between $\fata$ and the data (this reduces the posterior density) but excellent agreement between $\fata$ and $\sfC$ (this increases the posterior density). The former factor is bounded, while the latter is unbounded as we approach $\sfC = 0$. Thus there is an asymptote: the posterior density becomes infinite at $\fata = 0, \sfC = 0$ (note though that this behaviour affects only a small volume in parameter space; the integrated probability at the asymptote is tiny). The asymptotic behaviour contrasts, e.g., with a gaussian distribution, where the density reaches a finite maximum (a `peak') within the interior of the parameter space. If we reparameterise to coordinates $\{\fata, \sfG = \ln(\sfC)\}$ then the same asymptotic behaviour occurs, only now at $\sfG$ near minus infinity. However when $\sfG$ is large negative the posterior density falls off very quickly when the $\fata$ move away from zero. Thus in these coordinates the posterior density shape will resemble a funnel, or a stingray (Fig.~\ref{sting}). Such an extreme shape is difficult to sample from using an HMC sampler. A remedy can be to find a different parameterisation, in order to avoid the occurrence of the stingray (or, as we do, to move it to a less bothersome location in parameter space). For alternative approaches, see \cite{2009-Jewell}, \cite{2016-Racine}, \cite{Millea2021}.

In an analysis of a single $a_{\ell m}$ and a single $C_\ell$, the ratio $x = a_{\ell m}/\sqrt{C_\ell}$ will always be of unit variance. Therefore, although it may not solve all problems, switching to $x$ can avoid the occurrence of a stingray for small $C_{\ell}$. As we analyze many multipoles and power spectra jointly, the multivariate analogue is to flatten all $\fata_{\ell m}$ modes of the same $\ell$ jointly. This can be achieved by Cholesky decomposing the covariance matrices $\sfC_\ell$.

Each matrix $\sfC_{\ell}$ ($\ell_{\textrm{min}} \leq \ell \leq \ell_{\textrm{max}}$) is relatively small (for example, dimension $3$ for a CMB $TEB$-analysis, and dimension $2N_{\rm b}$ for a cosmic shear analysis with $N_{\rm b}$ redshift bins and both $E$- and $B$-modes) and hence can be decomposed at reasonable numerical cost.

In the rest of this section we fix $\ell$ (between $\ell_{\rm min}$ and $\ell_{\rm max}$) and for notational simplicity we drop the dependence on $\ell$ (so for example $\fata$ now refers to $a_{\ell m}$ for varying $m$ and this fixed $\ell$, while $\sfC$ refers to what was previously called $\sfC_\ell$).

Each matrix $\sfC$ is relatively small (for example, dimension $3$ for a CMB $TEB$-analysis, and dimension $2N_{\rm b}$ for a cosmic shear analysis with $N_{\rm b}$ redshift bins and both $E$- and $B$-modes) and hence can be decomposed at reasonable numerical cost.

For positive definite $\sfC$ the Cholesky factor $\sfL$ is the unique lower-triangular matrix with positive elements on the diagonal satisfying $\sfC = \sfL\sfL^{\transpose{}}$. Given $\fata$ define $\fatx$ via
\begin{equation}
    \fatx = \sfL^{-1}\fata;
    \label{Eq:Scale_fata}
\end{equation}
then if $\fata$ has mean zero and covariance $\sfC$ then $\fatx$ will have mean zero and unit covariance
\begin{equation}
\langle \fatx \fatx^{\transpose{}} \rangle = \mathbb{I}.
\end{equation}
We define the \textit{diagonal-log} $\sfK$ of $\sfL$ to be 
\begin{equation}
\sfK_{\alpha \beta} = 
\left\{
	\begin{array}{ll}
		\ln(\sfL_{\alpha \beta})  & \mbox{if } \alpha = \beta\, , \\
		\sfL_{\alpha \beta} & \mbox{otherwise.}
	\end{array}
\right.
\label{Eq:DiagonalLog_L}
\end{equation}
Then $\{ \fatx, \sfK \}$ provides an alternative parameterisation of $\{ \fata, \sfC \}$; note that an arbitrary $\sfK$ will generate a positive-definite $\sfC$.

\subsection{Jacobian for Cholesky coordinates}

Reparameterising from $\{ \fata, \sfC \}$ to $\{ \fatx, \sfK \}$ introduces a Jacobian factor in the posterior density. Now \begin{equation}
 p(\fatx,\sfK) \ \mathd \fatx \ \mathd \sfK = p(\fata,\sfC) \ \mathd \fata \ \mathd \sfC 
\end{equation}
and so the Jacobian factor is
\begin{equation}
    | \sfJ | = \begin{vmatrix} \partial \fata/\partial \fatx & \partial \fata/\partial \sfK\\
    \partial \sfC/\partial \fatx & \partial \sfC/\partial \sfK\\
    \end{vmatrix} = | \partial \fata/\partial \fatx | \ | \partial \sfC/\partial \sfK |;
\end{equation}
the second equality follows from $ {\partial \sfC} / {\partial \fatx} = 0 $ (since it is evaluated at constant $\sfK$).
We have $\fata = \sfL \fatx$ and so 
\begin{equation}
| \partial \fata/\partial \fatx | = |\sfL| = \prod_{\alpha=1}^{n} \sfL_{\alpha \alpha}.
\end{equation}

The determinant of the transformation from $\sfC$ to $\sfL$ is \citep{RedMatrixBook}:
\begin{equation}
| \partial \sfC/\partial \sfL | = 2^n\prod_{\alpha=1}^n \sfL_{\alpha \alpha}^{n+1-\alpha},
\end{equation}
while
\begin{equation}
| \partial \sfL/\partial \sfK | = \prod_{\alpha=1}^{n} \frac{\partial \sfL_{\alpha \alpha}}{\partial \sfK_{\alpha \alpha}} = \prod_{\alpha=1}^{n} \sfL_{\alpha \alpha}.
\end{equation}

Combining these results yields
\begin{equation}
    | \sfJ | = 2^n \prod_{\alpha=1}^n \sfL_{\alpha \alpha} ^{n+3-\alpha} = 2^n \exp \left[ \sum_{\alpha=1}^n (n+3-\alpha)\sfK_{\alpha \alpha} \right] \ .
    \label{Eq:Jax}
\end{equation}

\subsection{Negative Log Posterior in Cholesky coordinates}
\label{SSec:nlp}

In Cholesky coordinates $\{ \fatx, \sfK \}$ the negative logarithm $\nlp{}$ of the posterior (Eq.~\eqref{Eq:Cond_Posterior}) becomes:
\begin{equation}
\begin{split}
\nlp{}(\fatx, \sfK) & = \half (\fatd - \sfY\sfL\fatx)^{\rm T}\sfN^{-1}(\fatd - \sfY\sfL\fatx) + \frac{\fatx^{\rm T} \fatx}{2} \\
& + \sum_{\alpha=1}^n (\alpha - 2 - n -2q) \sfK_{\alpha \alpha} \ .
\end{split}
\end{equation}
Here we have included the Jacobian term (Eq.~\eqref{Eq:Jax}), we have used for example $\fata = \sfL\fatx$ and $|\sfC| = |\sfL|^2$, and we have silently dropped irrelevant additive constants. The final summand includes an implied sum over $\ell$.

This new coordinate system creates a posterior shape that can be explored more efficiently by a HMC sampler than the shape of the posterior in the $\{\fata, \ln(\sfC)\}$ coordinates. The data products outputted by the sampler are transformed back to the usual $\{\fata, \sfC \}$-parameters that a cosmologist would expect.

\subsection{Gradients in Cholesky coordinates}
\label{SSec:gradients}

The HMC sampler requires for its leapfrog routine the derivatives of the potential (i.e., the negative logarithm of the posterior) $\nlp{}$ with respect to both $\fatx$ and $\sfK$ parameters. Straightforward calculations show that the derivative with respect to $\fatx$ is 
\begin{equation}
\frac{\partial\nlp{}}{\partial \fatx_\alpha}  = -[\sfL^{\transpose{}} \sfY^{\transpose{}} \sfN^{-1}(\fatd-\sfY\sfL\fatx)]_\alpha +\fatx_\alpha \,
\end{equation}
and that the derivative with respect to $\sfK$ is 
\begin{equation}
\begin{split}
    \frac{\partial\nlp{}}{\partial \sfK_{\alpha\beta}} = & \Big( - (\delta_{\alpha \beta}^K (\sfL_{\alpha \beta} - 1) + 1 ) \\
    & \, \times \fatx_\beta[\sfY^{\transpose{}} \sfN^{-1}(\fatd-\sfY\sfL\fatx)]_{\alpha} \Big) \\
    & \, + \delta_{\alpha \beta}^K (\alpha-2-n-2q) \ ;
\end{split}
\end{equation}
here $\delta^K$ is the Kronecker delta.

\subsection{Hessian in Cholesky coordinates}
The Hessian of the posterior can be a good approximation for the mass matrix of an HMC sampler. If one does not want to (or cannot) use the full Hessian, one could still set the leapfrog step size to be the inverse square root of the diagonal of the Hessian; \almanac{} sets the initial step sizes in this way (although the step sizes are later tuned to better values). The diagonal elements of the Hessian are
\begin{equation} \label{Eq:hessian_x}
    \frac{\partial^2\nlp{}}{\partial \fatx_{\alpha} \partial \fatx_{\alpha}} = [\sfL^{\transpose{}} \sfY^{\transpose{}} \sfN^{-1} \sfY \sfL]_{\alpha \alpha} + 1
\end{equation}
and
\begin{equation} \label{Eq:hessian_K}
\begin{split}
    \frac{\partial^2\nlp{}}{\partial \sfK_{\alpha\beta} \partial \sfK_{\alpha\beta}} = & (\delta_{\alpha \beta}^K (\sfL_{\alpha \beta}^2 - 1) + 1 ) \fatx_\beta^2 [\sfY^{\transpose{}} \sfN^{-1} \sfY]_{\alpha \alpha}  \\
    & \, - \delta_{\alpha \beta}^K \sfL_{\alpha \beta} \fatx_\beta [\sfY^{\transpose{}} \sfN^{-1}(\fatd-\sfY\sfL\fatx)]_{\alpha} \ .
\end{split}
\end{equation}

We may estimate the diagonal terms of $[\sfY^{\transpose{}} \sfN^{-1} \sfY]_{\alpha \alpha}$ using Monte Carlo integration, as described in the Appendix of \citeCore{}. This technique allows us to calculate Eq.~\ref{Eq:hessian_x} and the first term in Eq.~\ref{Eq:hessian_K}; however, the second term in Eq.~\ref{Eq:hessian_K} is challenging to calculate efficiently and we therefore estimate it by numerical differentiation.

We have suppressed in this outline some details that are required for practical implementation. The values in $\sfL$, when applied to $\fatx$, must be used repeatedly ($2\ell+1$ times); this introduces $2\ell+1$ factors into the log determinant of the Jacobian (which then cancels out of the negative log posterior). It also introduces a sum over $2\ell+1$ terms in the $\sfK$ gradient, and a double sum in the $\sfK$ hessian. We must also account for the variance being split between the real and imaginary parts of $a_{\ell m}$ except when $m=0$; this introduces factors of $\sqrt{2}/2$ when $m \neq 0$.

\subsection{Reduced shear}

The quantity that controls the shape distortion is not strictly the shear, but rather the reduced shear \citep[e.g.][]{2017-Bartelmann}, $\gamma/(1-\kappa)$. For the tests in this paper we have not included reduced shear. In the future we intend to include reduced shear in the posterior calculation. This will complicate the calculation of the derivatives and we propose not including reduced shear in the derivatives calculation. The small (percent-level) errors in the resulting trajectories will then be dealt with completely by the Metropolis-Hastings acceptance/rejection step, which also deals with inaccuracies in the leapfrog integrator.

\section{Simulation}\label{Sec:Sims}

\begin{figure*}
    \centering 
    \includegraphics[width=\textwidth]{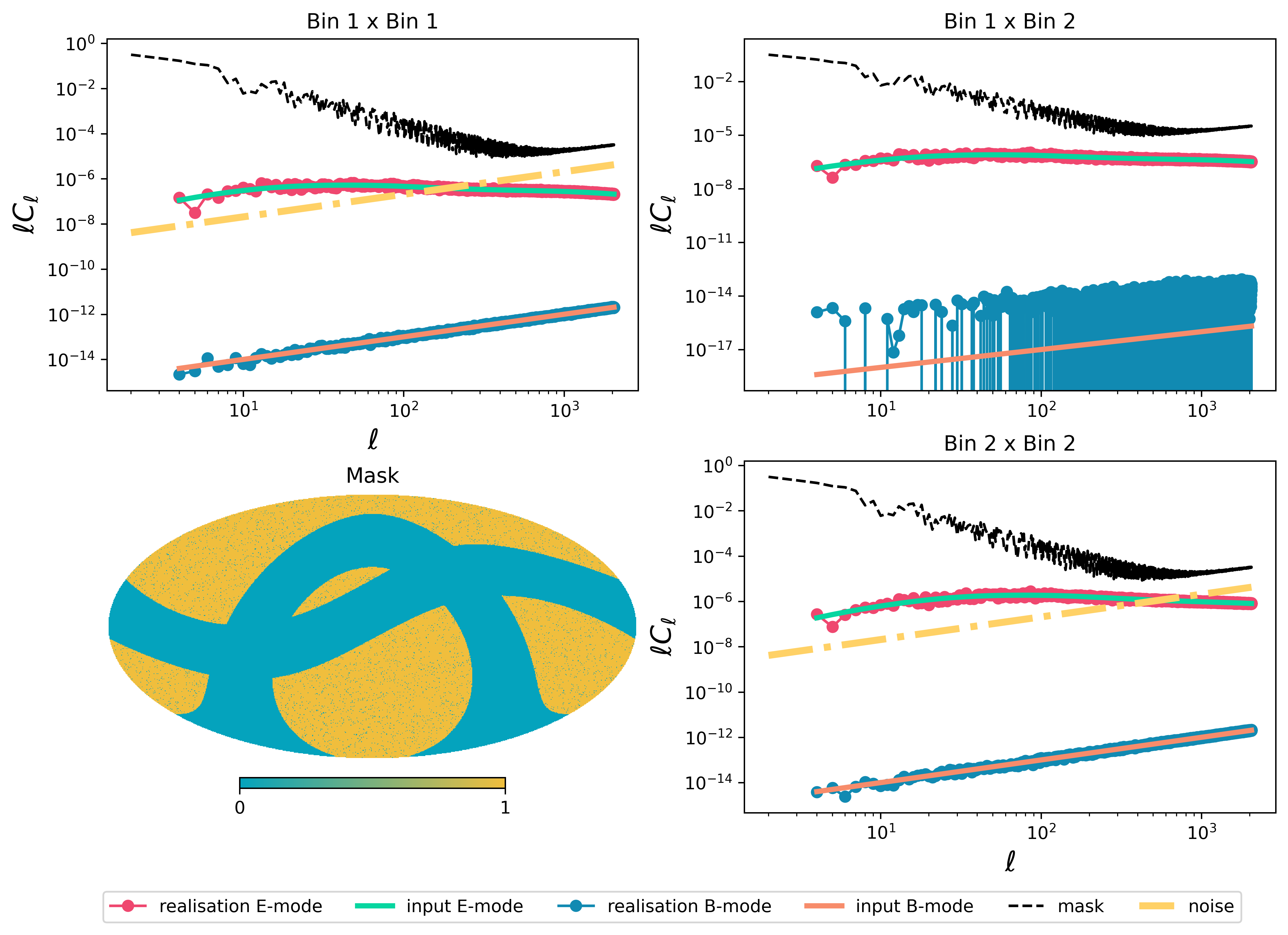}
    \caption{Setup for the Euclid-like cosmic shear survey simulation, showing the input angular power spectra to the GRF simulation for the $E$-modes (green line) and $B$-modes (orange line), the mask angular power spectra (dashed black line), the noise power spectra (yellow dot-dashed line), and the data realisation angular power spectra for $E$- and $B$-modes (pink and blue dots, respectively). The bottom left figure shows the angular mask used in the simulation.}
    \label{Fig:Simulation}
\end{figure*}
In order to demonstrate \almanac{}'s application to the next generation of cosmic shear experiments, we used gaussian random fields (GRF, \citealt{1986-Bardeen,1992-Hoffman}) to simulate a Euclid-like cosmic shear survey. Euclid is an excellent showcase for \almanac{}'s performance and application due to sky coverage, complex mask geometry, and expected data quality. Since \almanac{} infers all scales available from a given multipole range to properly forward-model the underlying field, galaxy surveys with large sky coverage such as Euclid and LSST are ideal data sets for the methodology described in this work.

The Euclid photometric sample will likely be divided into ten redshift tomographic bins with an equal effective galaxy number density of $\bar{n} = 3$ galaxies/arcmin$^2$ for each bin \citep{Euclid,2020-EuclidForecastAll,2020-TutusausEuclid,2021-EuclidTaylor,2022-EuclidWide}. We arbitrarily select two of these ten tomographic bins ($0.65 \leq z < 0.79$ and $1.54 \leq z < 1.83$)
to define the weak lensing tracers for the underlying theoretical angular power spectra. The fiducial $\sfC_{\ell}$ for the $E$-modes used in the simulation were calculated using \textsc{CCL} \citep{2019-CCL} with a Planck 2018 fiducial cosmology \citep{Planck2018}\footnote{$\Omega_{\rm cdm} = 0.27$; $\Omega_{\rm b} = 0.045$; $\Omega_{\rm k} = 0$; $n_{\rm s} = 0.96$; $\sigma_8 = 0.83$; $h=0.67$.}, in a range between $\ell_{\rm min} = 4$ and $\ell_{\max} = 2048$. Meanwhile, the $B$-mode power spectra were set to $10^{-15}$ and the $\sfC_{\ell}^{E_iB_j}$ were set to $10^{-20}$, both constant across the multipole range used for the $E$-mode power spectra.

The survey angular geometry mask was created by masking out the galactic plane and the ecliptic based on the description in \cite{2022-EuclidWide}. We then randomly masked pixels using a Poisson distribution to mimic the effects of stars in the field. The resulting mask is shown in the bottom left of Fig.~\ref{Fig:Simulation}.

The last element in the simulation is shape noise, taken to be $\sigma_{\epsilon} = 0.27$. Note (see Fig.~\ref{Fig:Simulation}) that the $B$-modes are almost six orders of magnitude below the noise power spectra.

The noise covariance in real (pixel) space $\sfN$ is taken to be a diagonal matrix. It has $2N_bN_{\rm pix}$ diagonal elements (the factor of $2$ corresponding to the two components of shear); for a given shear component, tomographic bin $i$, and pixel $p$ the corresponding entry in $\sfN$ is ${\sigma_{\epsilon}^2} / {(2\bar{n}_i)}$ if $p$ is unmasked and infinity otherwise. Here $\bar{n}_i$ is the lensing source number per pixel in bin $i$.

The simulations are performed using a \textsc{HEALPix} \citep{1999-Healpix} resolution of $N_{\rm side} = 1024$  ($\Delta\Omega_{\rm pix} = 11.8$ arcmin$^2$) by sampling correlated $\fata$ from a multivariate gaussian distribution, $\fata \sim \calN(0; \sfC_{\ell})$, and then transforming the signal maps into real space. Next, using the aforementioned galaxy density and shape noise, we add gaussian noise to the signal maps with a variance of $\sigma^2 = \sigma_{\epsilon}^2/(2\bar{n}_i) $ per component. In a final step, we apply the mask shown in Fig.~\ref{Fig:Simulation} to all the galaxy ellipticity data maps. 

Although a more realistic N-body simulation with ray-tracing would include higher-order effects into the tests, a GRF simulation allows for more control over the quantities we are probing and testing. It lets us quickly simulate correlated tomographic cosmic shear data, allowing full control and understanding of the effect of signal, noise, and mask; a GRF is the most reliable way to have full control of the simulation's underlying angular power spectrum and noise. This approach also allows us to keep fixed the simulation's signal realisation while turning off shape noise and mask effects to produce the ground truth. More complex simulations should not pose any difficulties for the \almanac{} analysis presented in Sec.~\ref{Sec:Results} as the method is sensitive to the statistical properties of the input data and makes no assumptions on the underlying physics of the analysed data maps beyond the spin, noise, and statistical isotropy.

\section{Results}\label{Sec:Results}

We apply the \almanac{} BHM described in Sec.~\ref{Sec:Methods} to the Euclid-like simulation described in Sec.~\ref{Sec:Sims} probing a multipole range $ 4 \le \ell \le 2048 $ with $N_{\rm side} = 1024$. Here we describe the main outputs from this analysis: the marginalised angular power spectrum and the inferred shear and convergence full sky maps. We further demonstrate other benefits of having the full joint posteriors of maps and angular power spectra, and we discuss implications for future cosmic shear surveys (such as our approach to $EB$ leakage, and the recovery of large scale modes).

We keep the $B$-mode auto-power spectra in the modelling, with a small signal amplitude in the simulated data; although in the standard model of cosmology these modes are expected to be negligible for cosmic shear, we want the ability to 
detect systematic contamination that could appear through the $B$-modes \citep{2006-Heymans, 2019-Asgari-BModes}.

The full posterior contains all the spin-2 spherical harmonics for both redshift tomographic bins and their angular power spectra; this results in a total of 16,813,990 dimensions (of which only 20,450 are for the angular power spectra, $\sfC_{\ell}$).

We ran two simultaneous chains using different starting points and different random seeds.
Each chain contains {$N_{\rm samples} \approx 2.25 \times 10^5$} samples (after burn-in and tuning) and each had an acceptance ratio of $\sim 86\%$. For both chains we used {$N_{\rm tune} = 2 \times 10^4$} and {$N_{\rm lf-tune} = 5 \times 10^3$} samples to perform the step size tuning described in \citeCore{}. Unless stated otherwise, all the results presented in this section come from combining the chains (once the chains were individually converged --- see Sec.~\ref{Sec:Conver} for a discussion of the convergence diagnostics for chains used in this analysis).

\subsection{Inferred Angular Power Spectra}
\begin{figure*}
    \centering 
    \includegraphics[width=0.95\textwidth]{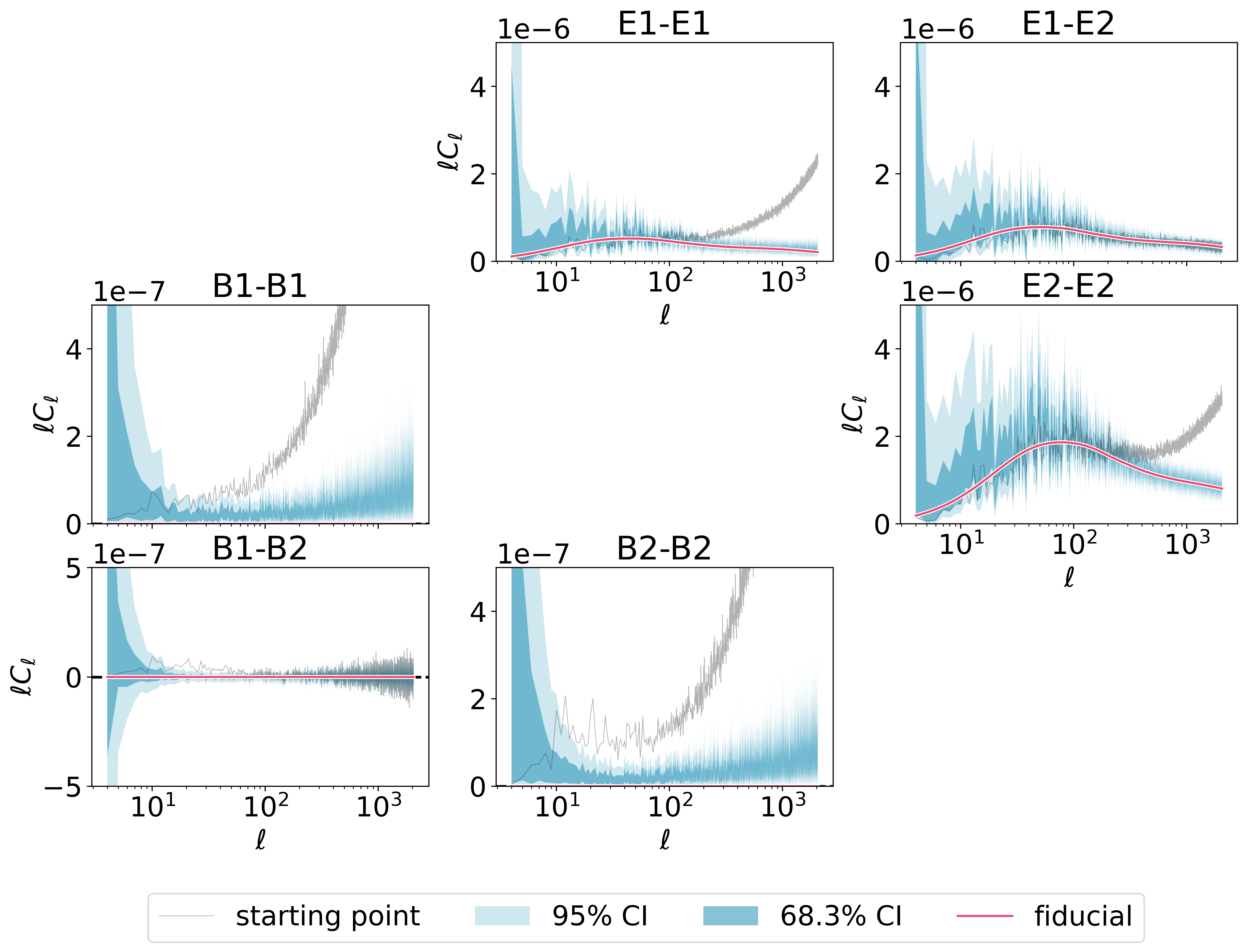}
    \caption{Inferred marginalised angular power spectra for the $E$- and $B$-modes. The blue (resp. light blue) hashed area is the 68.3\% (resp. 95\%) C.I. The starting point for one of the chains is shown in light grey. The red line is the ground truth given to the input simulation shown in Fig.~\ref{Fig:Simulation}. The credible intervals for the $B$-mode auto-power spectra extend close to the lower bound of zero.}  
    \label{Fig:Cls_E_B}
\end{figure*}

\begin{figure}
    \centering 
    \includegraphics[width=0.49\textwidth]{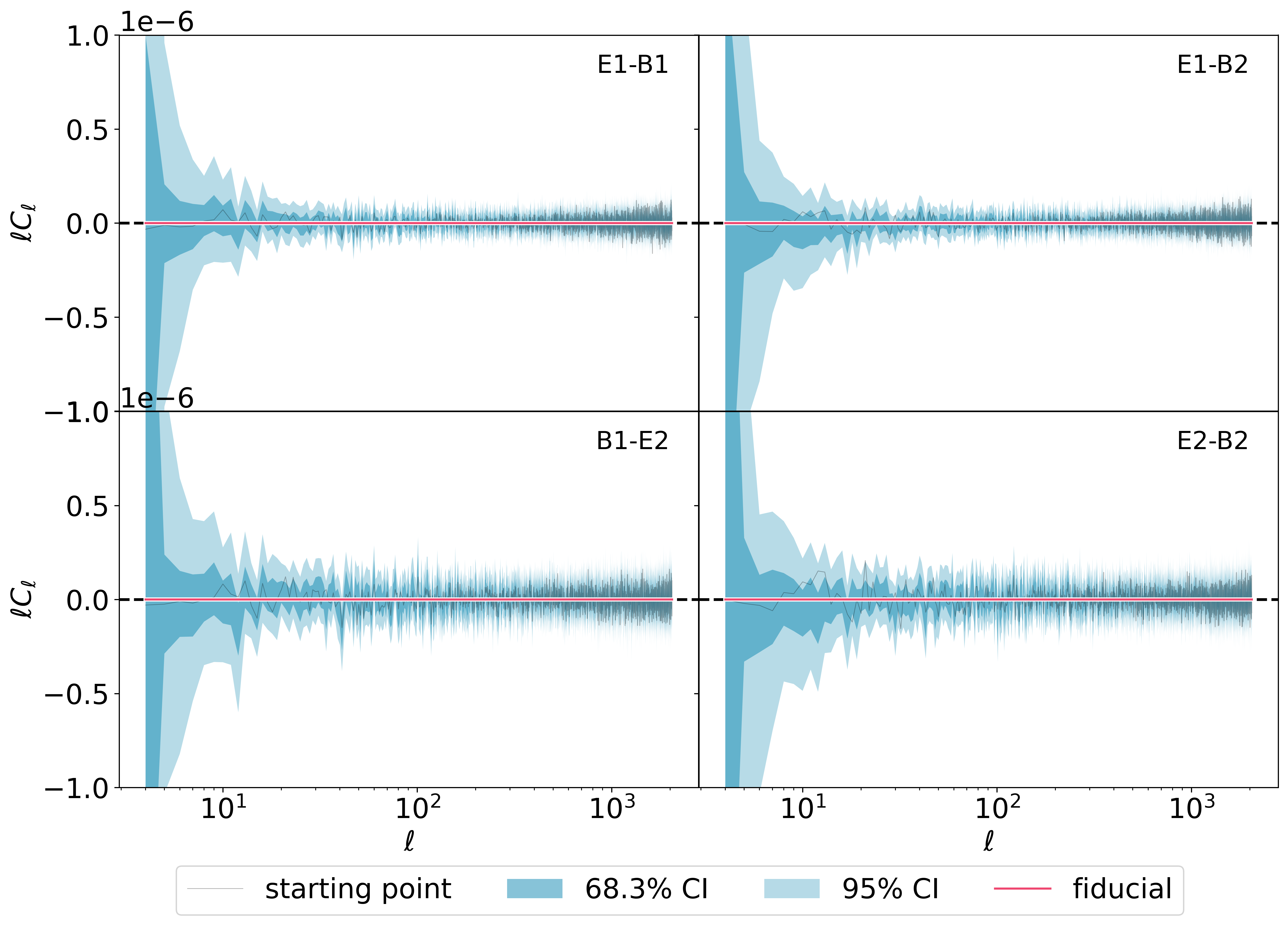}
    \caption{Similar to Fig.~\ref{Fig:Cls_E_B} but showing the parity violating modes instead.}
    \label{Fig:Cls_ParityViol}
\end{figure}

\begin{figure*}
    \centering 
    \includegraphics[width=\textwidth]{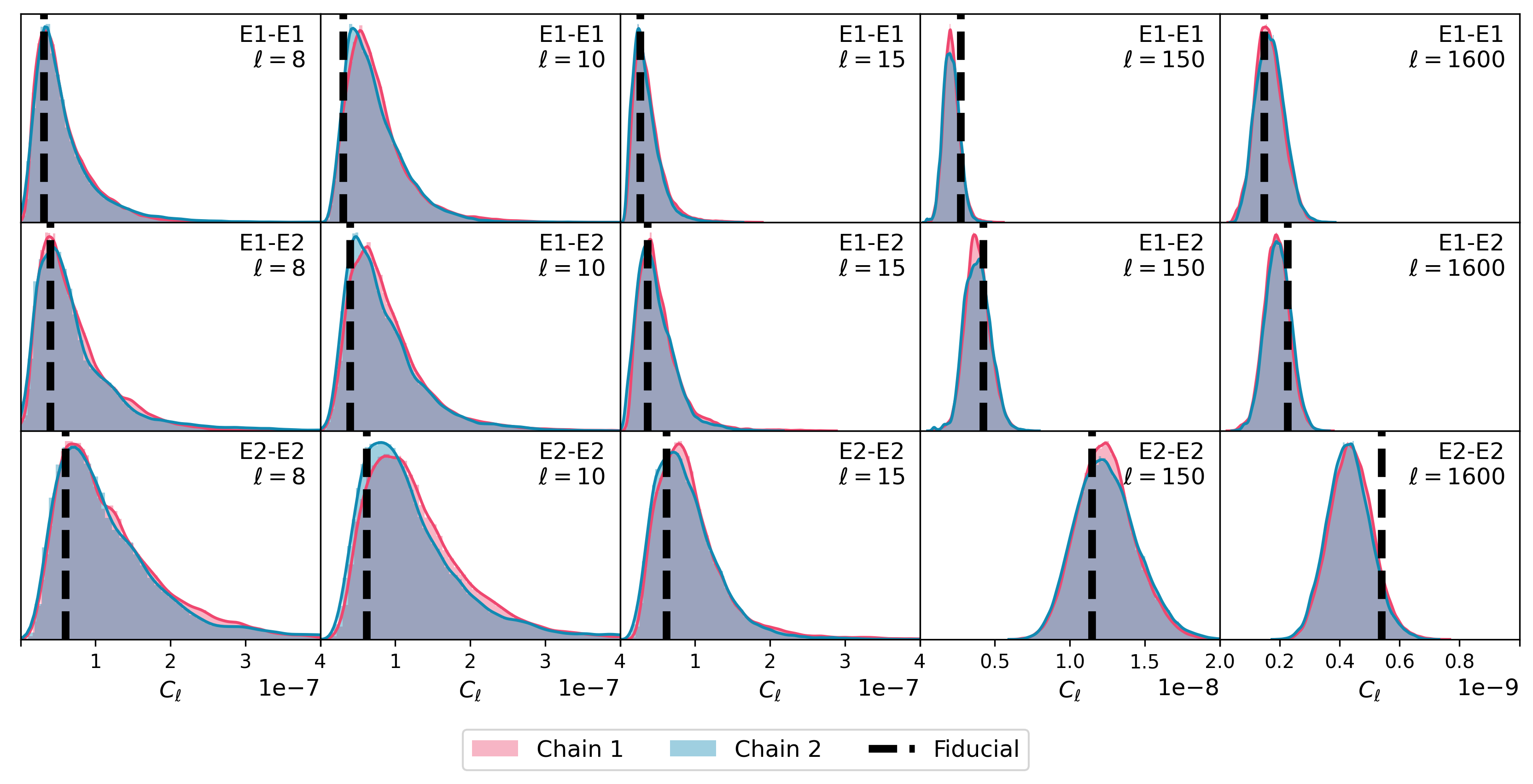}
    \caption{Marginalised one-dimensional posterior distributions for $E$-modes of different multipoles ($\ell = 8,\, 10,\, 15,\, 150,\, 1600$). As expected, marginalised posteriors for large scale modes are significantly non-gaussian while those for small scale modes are much closer to gaussian (due to the higher number of spherical harmonic coefficients available).}
    \label{Fig:EMode_Hist}
\end{figure*}

For each element of the power spectra we calculate the one-dimensional marginalised 68.3\% and 95\% credible intervals (C.I.) from the sampled posterior. The results are shown in Fig.~\ref{Fig:Cls_E_B} for the $E$- and $B$-modes in both tomographic bins and in Fig.~\ref{Fig:Cls_ParityViol} for the parity violating modes. The natural outcome of the analysis is the joint posterior of the power spectra and the maps. This high-dimensional distribution is not simple in shape, so we do not show point estimates (e.g., mean, mode), as they do not necessarily act as useful summaries of the posterior. The apparent noisy spikiness in the marginalised angular power spectra shown in Figs.~\ref{Fig:Cls_E_B} and \ref{Fig:Cls_ParityViol} is expected since the marginalised multipoles are independent and since we are not binning our results in bandpowers.

Fig.~\ref{Fig:Cls_E_B} demonstrates \almanac{}'s ability to infer the underlying angular power spectrum for cosmic shear data on the sphere up to scales relevant for the current and next generation of surveys. Although not a requirement for a Bayesian analysis, the marginalised $E$-mode $\sfC_{\ell}$s agree well with the input power spectra given to the Euclid-like simulations within the 68.3\% C.I. for all multipoles in the analysed range, including the largest scales available in the simulation. Our prior for $\sfC$ excludes non-positive-definite matrices, and hence all diagonal elements of $\sfC$ must be positive; in particular, the $B$-mode auto-power spectra cannot be zero or negative. The low signal-to-noise ratio of the B-mode observations implies that the marginalised $C_\ell^{BB}$ posterior will often peak at positive values even when no such signal is present. This is evident from the form of the map-marginalised posterior \citep{bjk98}. Moreover, sampling near the $C_\ell^{BB}=0$ boundary is challenging due to the funnel shape of the posterior in the full $(a_{\ell m}, C_\ell)$ parameter space. This region comprises a very small fraction of the full parameter volume. In most cases, the credible intervals extend to or very near the lower bound. Both the $B$-mode upper bound and the parity-violating modes are efficient probes for systematic contamination that can be calculated in a fully Bayesian way with the measurements produced by \almanac{}. 

We show several examples of marginalised one-dimensional posterior distributions of $E$-mode multipoles in Fig.~\ref{Fig:EMode_Hist}. This plot illustrates that the method can properly infer scales that would usually be considered too large (given the fraction of the sky made available by the survey's mask), since the data in the unmasked regions, in conjunction with the prior, constrain modes on all scales. 

\begin{figure}
    \centering 
    \includegraphics[width=0.49\textwidth]{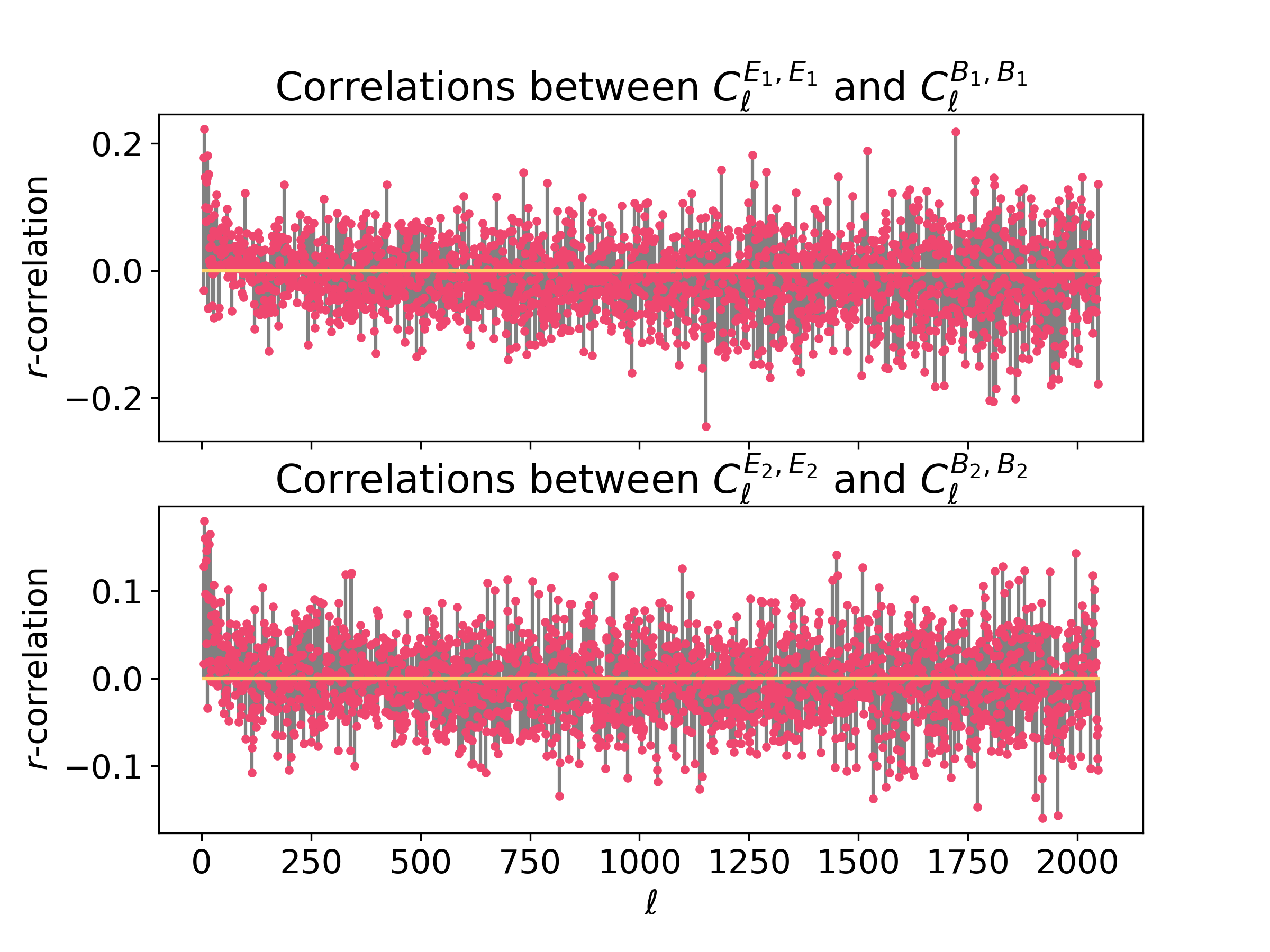}
    \caption{Correlations between the samples of same multipoles between auto-power spectra of $E$- and $B$-modes. For each pair of $C_{\ell}^{EE}$ and $C_{\ell}^{BB}$
    highly anti-correlated modes indicate that lower (higher) $B$-modes are degenerate with higher (lower) $E$-modes. We find no evidence for this effect in our analysis.}
    \label{Fig:EBLeakCorr}
\end{figure}

A particular issue when using point estimators of two-point statistics for weak lensing data is the occurrence of ambiguous modes due to the survey's mask and anisotropic noise. Commonly referred to as `$EB$ leakage' \citep{2003Bunn-EBLeak,2008-Bunn-EBLeak,2017-Leistedt,2019-Liu,2021-Nicola}, this effect causes high (low) $B$-modes to be confused with low (high) $E$-modes, yielding a negative correlation between these modes on the marginalised posterior distribution for a given multipole. Since we are not working with estimators of the $E$- and $B$-mode power spectra, instead inferring the full posterior of the underlying whole sky fields and their angular power spectra, such leakage of power from $E$ to $B$ does not arise. What may generically happen is that the posterior samples of $E$- and $B$-modes are correlated.
To check for this effect, we calculate the $r$-correlation factor for each pair of auto power spectra samples $\{ C^{E_i,E_i}_{\ell=L}, C^{B_i,B_i}_{\ell=L}\}$ for a given multipole $L$. The $r$-correlation factors for both tomographic bins are plotted in Fig.~\ref{Fig:EBLeakCorr}, which shows a rather small level of correlation.
Although we find no significant correlations between $E$- and $B$-modes in our analysis, we outline here another advantage of a Bayesian approach. Since we have the full posterior of angular power spectra, modes that demonstrate significant correlation can be removed by marginalising. We highlight, however, that since the full posterior distribution is known, this marginalisation would only be necessary if one were to take point estimates of the marginalised individual modes of the $E$-mode angular power spectra.

\subsection{Inferred Shear and Convergence Maps}\label{SSec:Maps}
\begin{figure*}
    \centering 
    \includegraphics[width=0.75\textwidth]{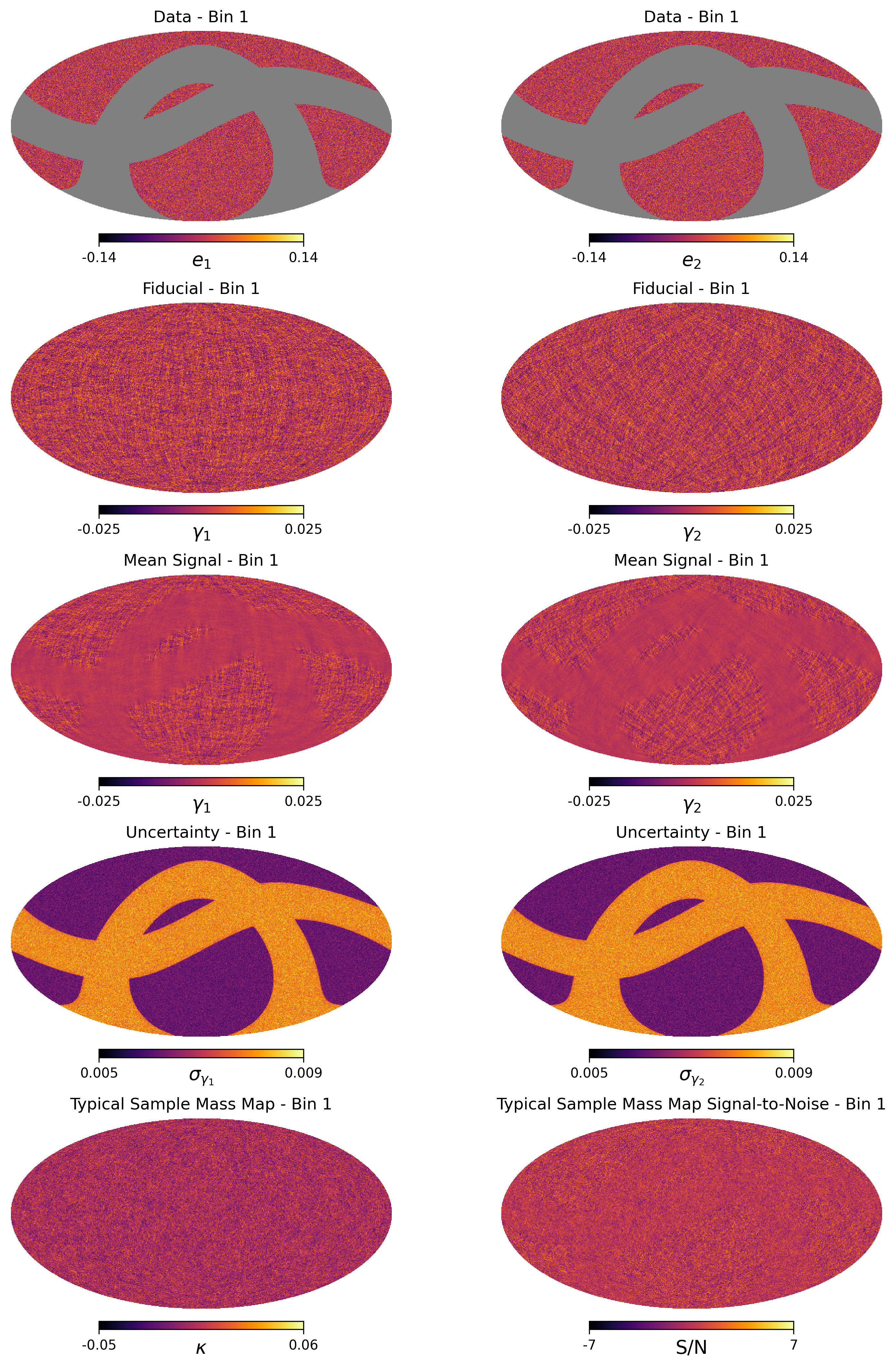}
    \caption{Maps for the first redshift tomographic bin ($0.65 \leq z < 0.79$). Row 1: simulated galaxy ellipticity data; row 2: its underlying cosmic shear signal (fiducial); row 3: recovered mean shear components; row 4: errors for the components of the recovered shear; row 5: typical convergence (left) and its signed S/N ratio (right). The fiducial maps contain the same signal realisation as the simulated maps but without the shape noise and without the mask. A `typical' map is a randomly selected (post burn-in) sample. The mean maps recovered most of the underlying structure shown in the fiducial maps in areas where data is present and are able to recover some of the large-scale information under the mask. The uncertainty on the recovered shear maps reflects the data missing due to the survey geometry.}
    \label{Fig:Maps_bins1}
\end{figure*}
\begin{figure*}
    \centering 
    \includegraphics[width=0.75\textwidth]{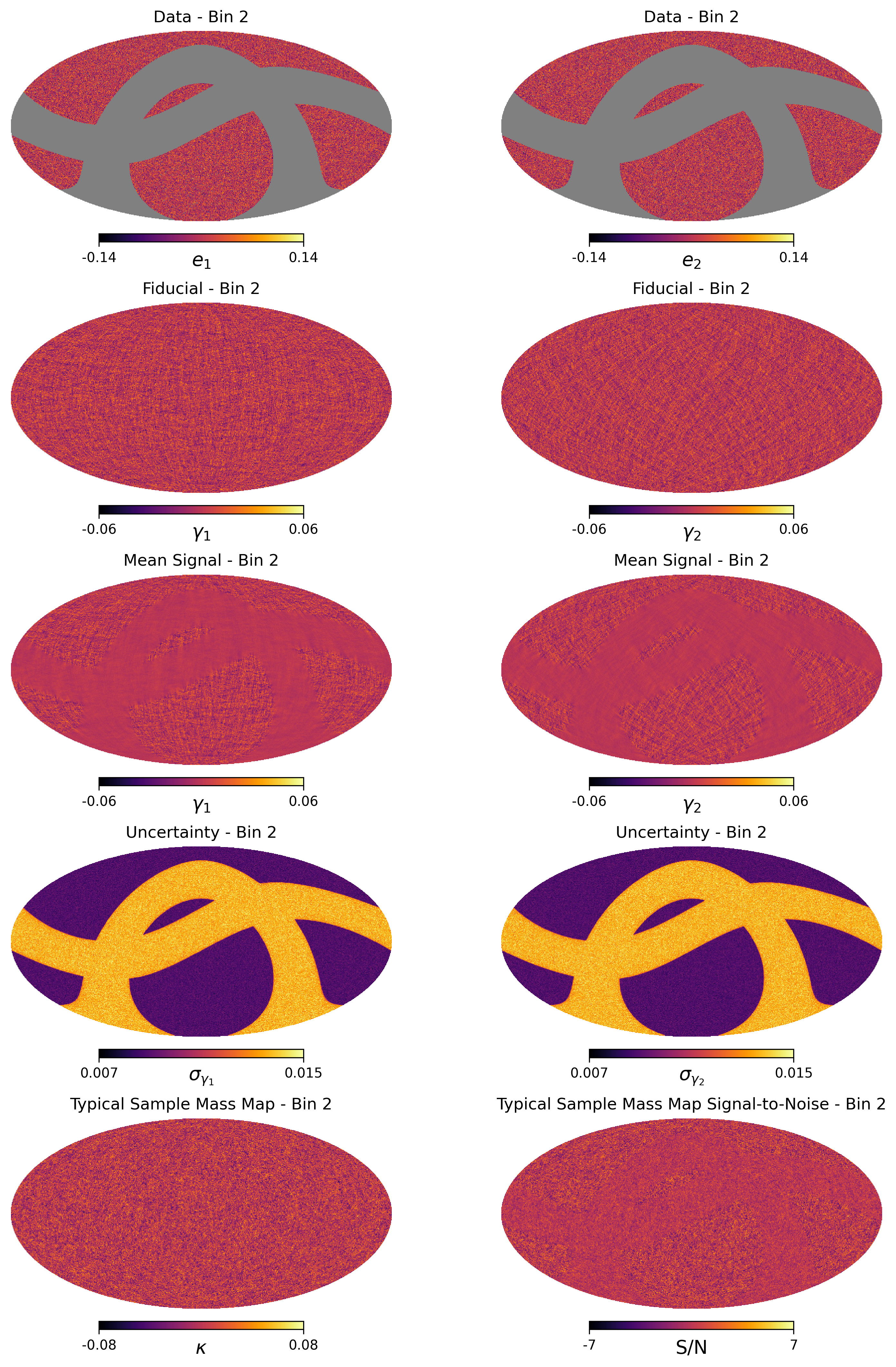}
    \caption{Same as Fig.~\ref{Fig:Maps_bins1} but for the second tomographic bin ($1.54 \leq z < 1.83$).}
    \label{Fig:Maps_bins2}
\end{figure*}

\begin{figure*}
    \centering 
    \includegraphics[width=\textwidth]{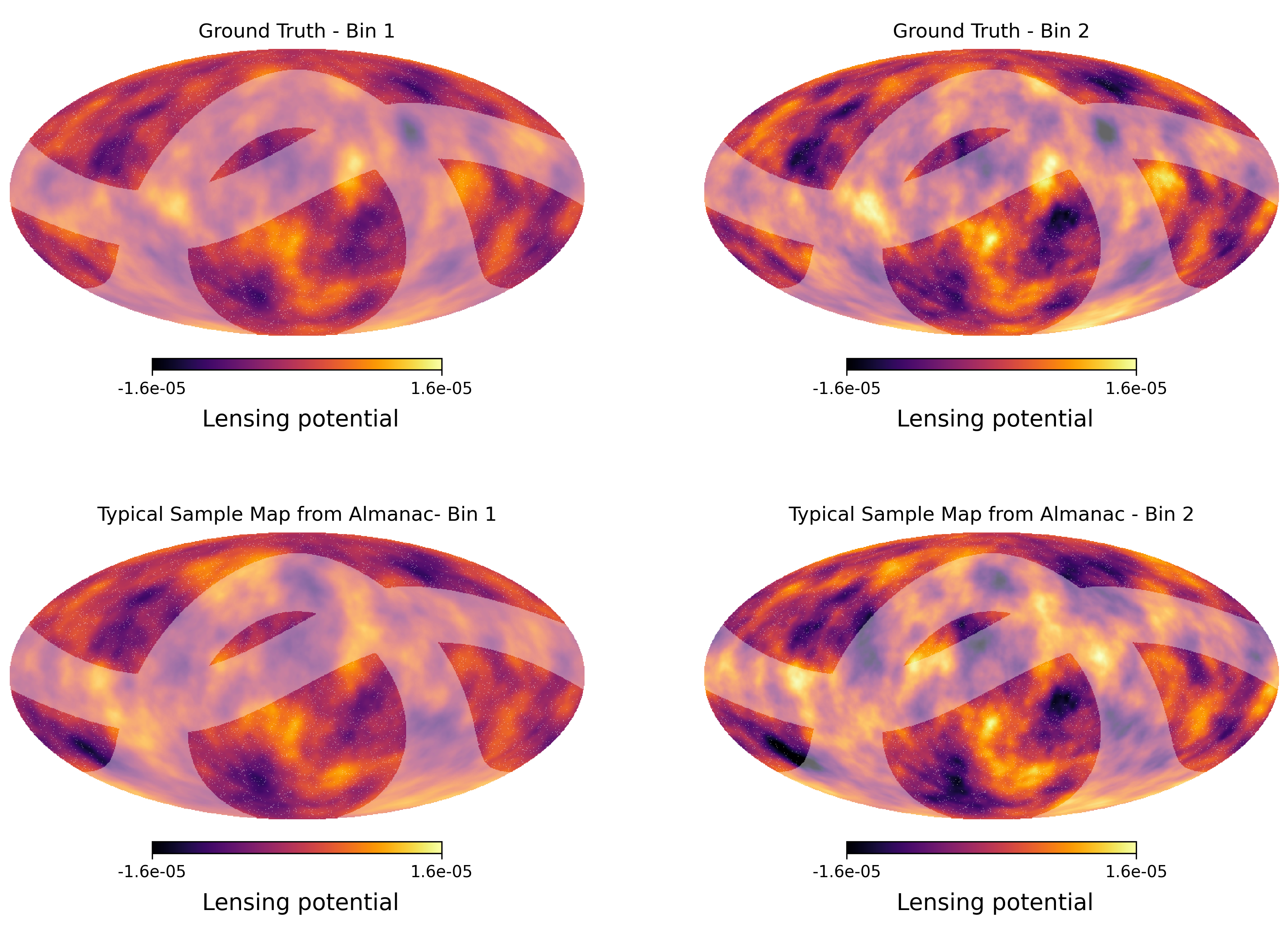}
    \caption{Reconstructed projected gravitational potential (or lensing potential), $\lpot{}(\hat{n})$, from typical \almanac{} maps for both tomographic bins. Top: the ground-truth lensing potential obtained from the fiducial maskless noiseless simulations (second row on Fig.~\ref{Fig:Maps_bins1} and \ref{Fig:Maps_bins2}). Bottom: projected lensing potential reconstructed from the simulated data (first row on Fig.~\ref{Fig:Maps_bins1} and \ref{Fig:Maps_bins2}) using a typical \almanac{} map. Here, we overlay the mask used in the simulations for visualisation purposes. }
    \label{Fig:LensingPotential}
\end{figure*}

\begin{figure}
    \centering 
    \includegraphics[width=0.47\textwidth]{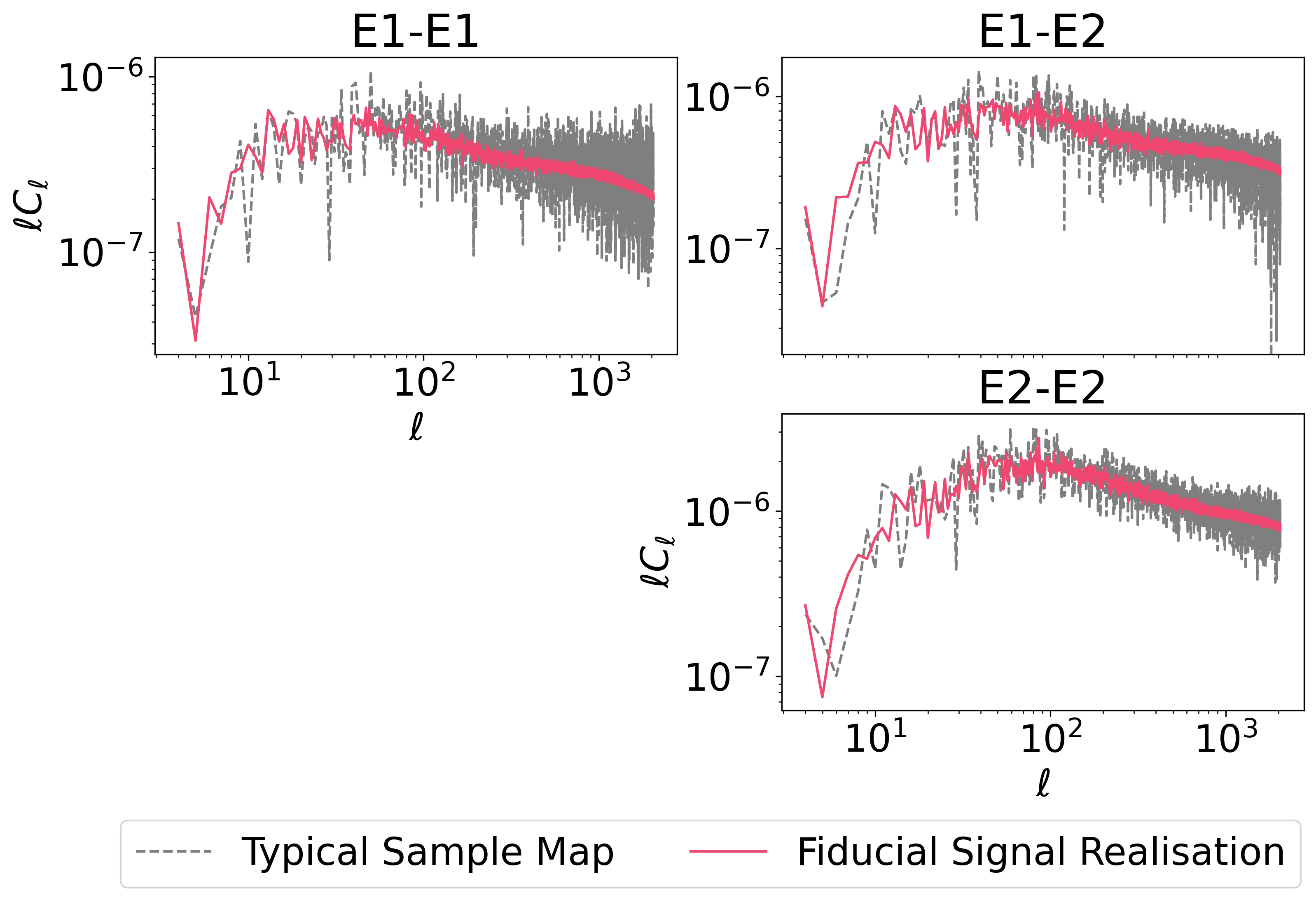}
    \caption{Consistency check between the inferred shear maps and the simulated signal map showing the $E$-mode angular power spectra for a typical map. In red we show the angular power spectra for the signal realisation in the simulations, while in grey we show the pseudo-$\sfC_{\ell}$ 
    for the typical map.}
    \label{Fig:RecovCls}
\end{figure}
\begin{figure}
    \centering 
    \includegraphics[width=0.47\textwidth]{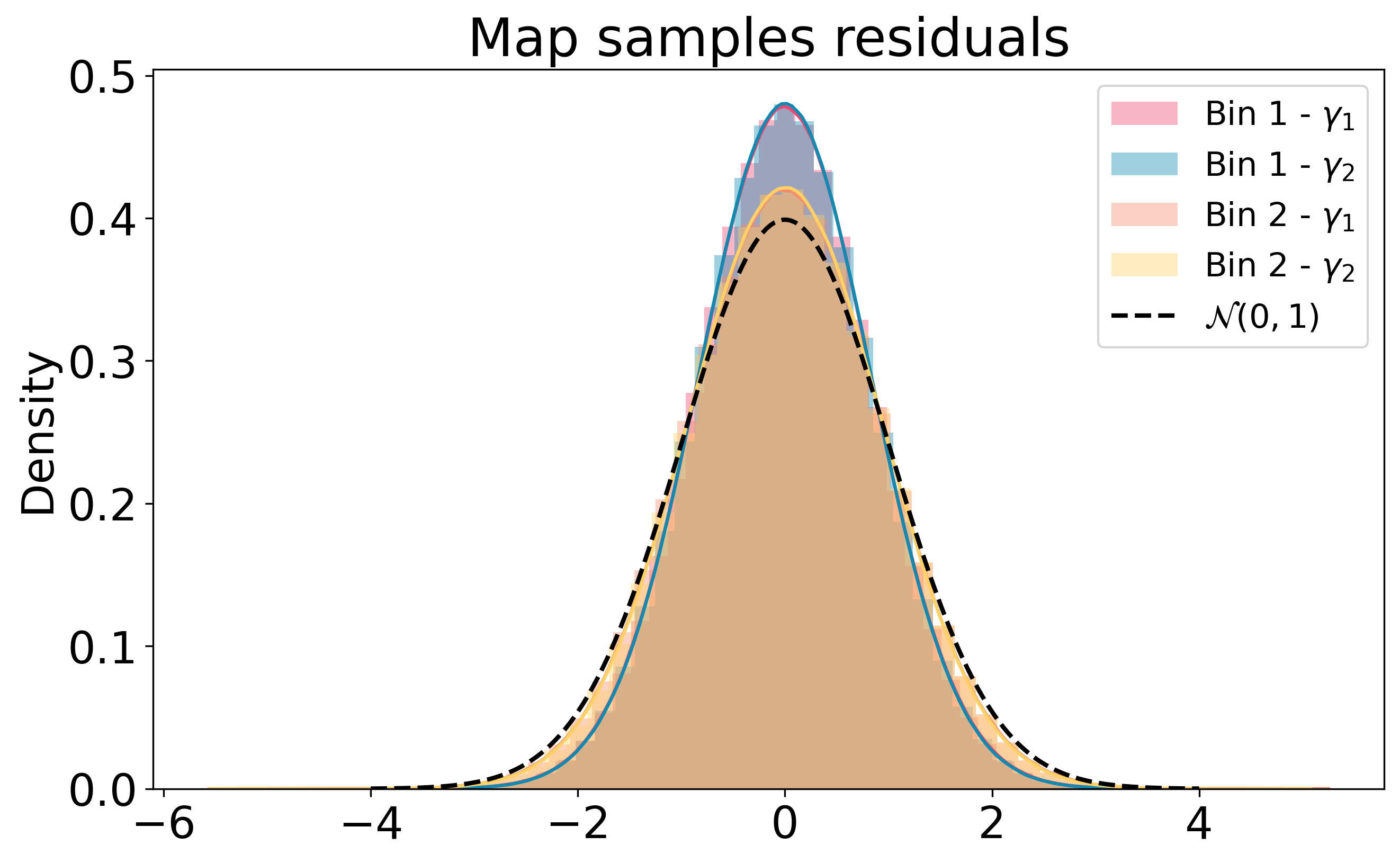}
    \caption{The residuals from the samples of the shear component maps obtained by \almanac{} compared to the underlying noiseless maskless signal map. The residuals are expected to be distributed as a standard normal (and we therefore show such a standard normal for comparison). }
    \label{Fig:MapResiduals}
\end{figure}
We next consider the spherical harmonic coefficients; these make up the larger subset of the joint set of parameters. In this section we analyse the posterior samples when marginalised over the angular power spectra. For each $\fata$ sample in the chains, we make a spherical harmonic transformation, following the formalism stated in Sec.~\ref{Sec:WLtheory}, to obtain real-space maps for shear and convergence. The $\fata$ samples are quite large, so to reduce file size (and hence increase speed) only one $\fata$ sample in 800 is saved; this thinning yields 577 samples across both chains.

The main results for the inferred shear and convergence maps and their uncertainties are shown in Fig.~\ref{Fig:Maps_bins1} (first bin, $0.65 \leq z < 0.79$) and Fig.~\ref{Fig:Maps_bins2} (second bin, $1.54 \leq z < 1.83$). In these figures, the data maps (first row) show the simulated galaxy ellipticities (here data is missing due to the survey geometry mask, and the maps contain shape noise according to the description in Sec.~\ref{Sec:Sims}). The fiducial maps (second row) show the same simulated maps but without the shape noise and survey mask (i.e., they contain only the underlying cosmic shear signal). The mean maps (third row) and standard deviation maps (fourth row) are calculated from the statistics of the $\fata$ samples transformed into real space. Finally, the `typical' convergence map (fifth row, left) is transformed from the $E^{\rm typical}_{\ell m}$ of a randomly selected (post burn-in) sample (compare Eq.~\eqref{Eq:gamma_lm_from_kappa}):
\begin{equation}
    \kappa^{\rm typical}_{\ell m} = -\sqrt{\frac{\ell(\ell+1)}{(\ell-1)(\ell+2)}}\ E^{\rm typical}_{\ell m}
\end{equation}
while the signed S/N ratio map (fifth row, right) is defined as
\begin{equation}
    (S/N)_{\kappa} \equiv \frac{\kappa^{\rm typical}(\theta,\phi)}{\sigma_{\kappa}(\theta,\phi)}
\end{equation}
where $\sigma_{\kappa}$ is the standard deviation of the convergence maps at the relevant pixel.

Comparing the mean shear maps to the fiducial map, we see that \almanac{} samples can properly recover the underlying structure of shears where the input galaxy ellipticity maps have observations. For the masked areas, where observations are not present, the mean maps display suppression of structure for small scales. This is expected from the averaging of the samples. Each sample is characterized by a power spectrum that contains small-scale modes, so the sampled maps have fine structures even in the masked regions. However, in the masked regions, the small-scale structure is not well constrained, so different samples have different structures there, and these tend to average out when the mean map is computed.
The masked regions also have higher posterior uncertainty, and this is reflected in the uncertainty maps shown in the fourth row of Figs.~\ref{Fig:Maps_bins1} and \ref{Fig:Maps_bins2}. 

Finally, we can use Eqs.~\eqref{Eq:Kappa_lm} and \eqref{Eq:gamma_lm} to obtain typical maps for the tomographic projected lensing potential (or the projected Newtonian gravitational potential, $\lpot{}$, as defined by Eq.~\eqref{Eq:lensing_potential}) for the full sky. These maps are shown in Fig.~\ref{Fig:LensingPotential}, where they are compared to the ground-truth obtained from the fiducial simulations (same underlying signal as the data but full-sky and noiseless). We overlay the mask (white semi-transparent region in the maps); this allows us to observe that \almanac{}'s inferred gravitational field is in excellent agreement with fiducial field in unmasked regions, while, as expected, the inferred map displays an overall similar structure under the mask but with different locations of voids and walls than that found in the fiducial map. 

We perform two main consistency checks in the recovered maps, one for the typical maps and another for the mean shear maps. The first test (see Fig.~\ref{Fig:RecovCls}) compares the underlying signal realisation from the fiducial maskless and noiseless simulation with the pseudo-$\sfC_{\ell}$ estimate of the typical shear maps. We see that it is a \textit{typical} map that best reflects the underlying two point statistics. As expected, the agreement between the two is better for the largest scales, while the typical map scatters around the fiducial value for the smaller scales (due to the missing information resulting from the mask). By contrast, the mean map will be over-smoothed in regions of low signal-to-noise ratio. One could also consider the statistics of the maximum a posteriori map, but the funnel shape of the posterior forces this to $a_{\ell m}=0$, and in fact the small parameter volume in the neighbourhood of the maximum means that even the maximum a posteriori \textit{sample} is far from this point.  The second test (see Fig.~\ref{Fig:MapResiduals}) checks the distribution of residuals between the shear maps samples and the fiducial shear map
\begin{equation}
    \mathcal{R}(\gamma_i) \equiv \frac{1}{\sigma_{\gamma_i}}\left( \bar{\gamma}_{i}  - \gamma^{\rm fiducial}_i\right)
\end{equation}
where $\bar{\gamma}_{i}$ is the mean map for the samples of shear component \textit{i}; these are expected to follow a standard normal distribution. The figure shows the distributions of the residuals for all recovered shear component maps, and a standard normal for comparison. The small difference seen for the lower redshift tomographic bins can be explained by their lower signal-to-noise affecting the ability of the method to accurately estimate these maps. However, the difference is below 1-$\sigma$ and does not raise concerns about the maps shown in Figs.~\ref{Fig:Maps_bins1} and \ref{Fig:Maps_bins2}.

\subsection{Convergence diagnostics}\label{Sec:Conver}

In this section we describe the convergence diagnostics used to assess the validity of the chains (and hence the validity of the results shown in the previous sections). Assessing convergence of a high-dimensional Monte Carlo chain is a notoriously difficult problem. We examined several convergence diagnostics on the individual chains, probing different aspects of the convergence: a trace of the log posterior, the Fraction of Missing Information, the correlation length of the chains, Hanson's statistic, and the Gelman-Rubin test. Further details are in \citeCore{}.

\begin{figure}%
    \centering
    \subfloat[\centering Log Posterior Trace]{{\includegraphics[width=.48\textwidth]{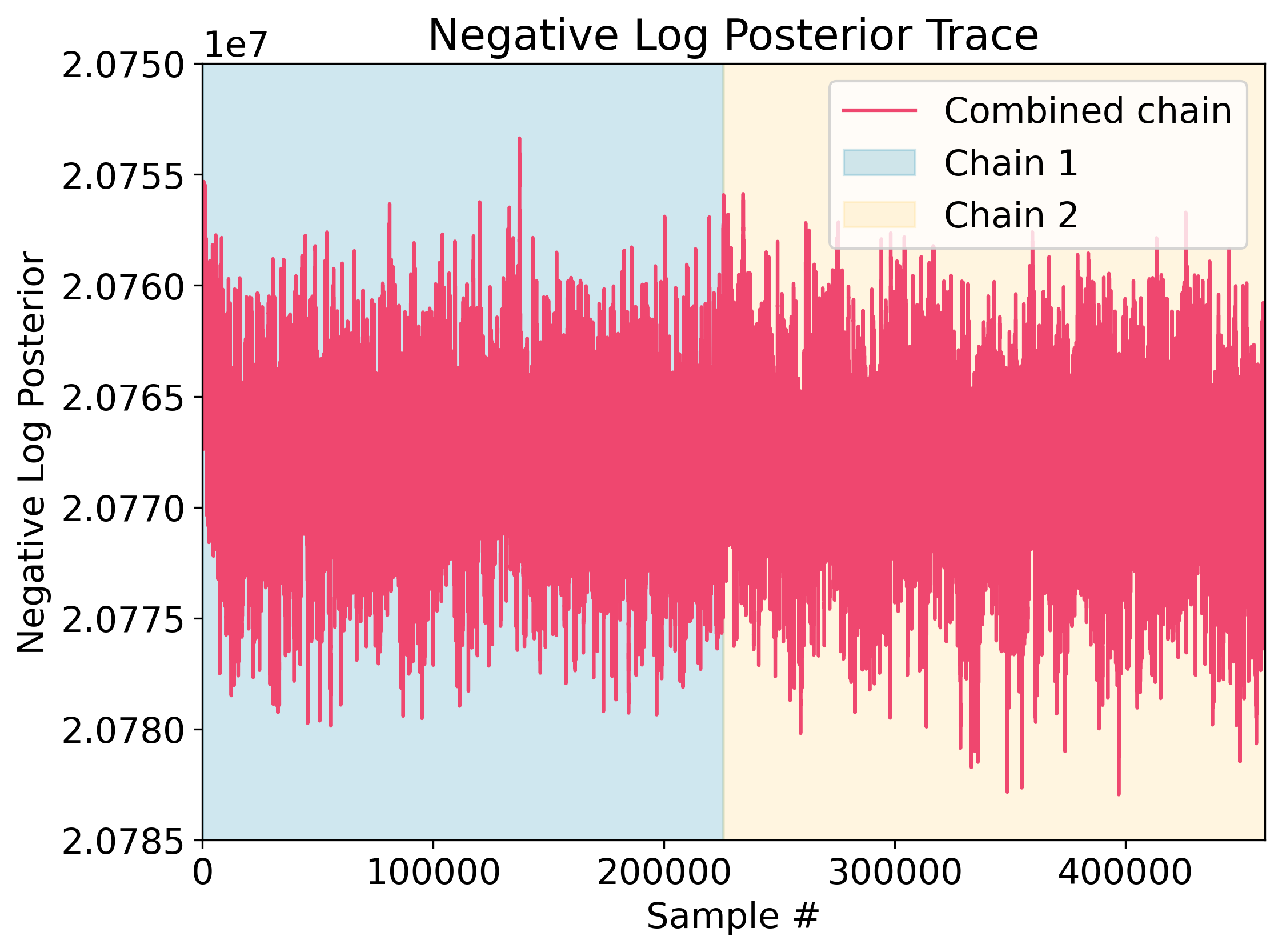} }}%
    \qquad
    \subfloat[\centering Log Posterior Histogram]{{\includegraphics[width=.50\textwidth]{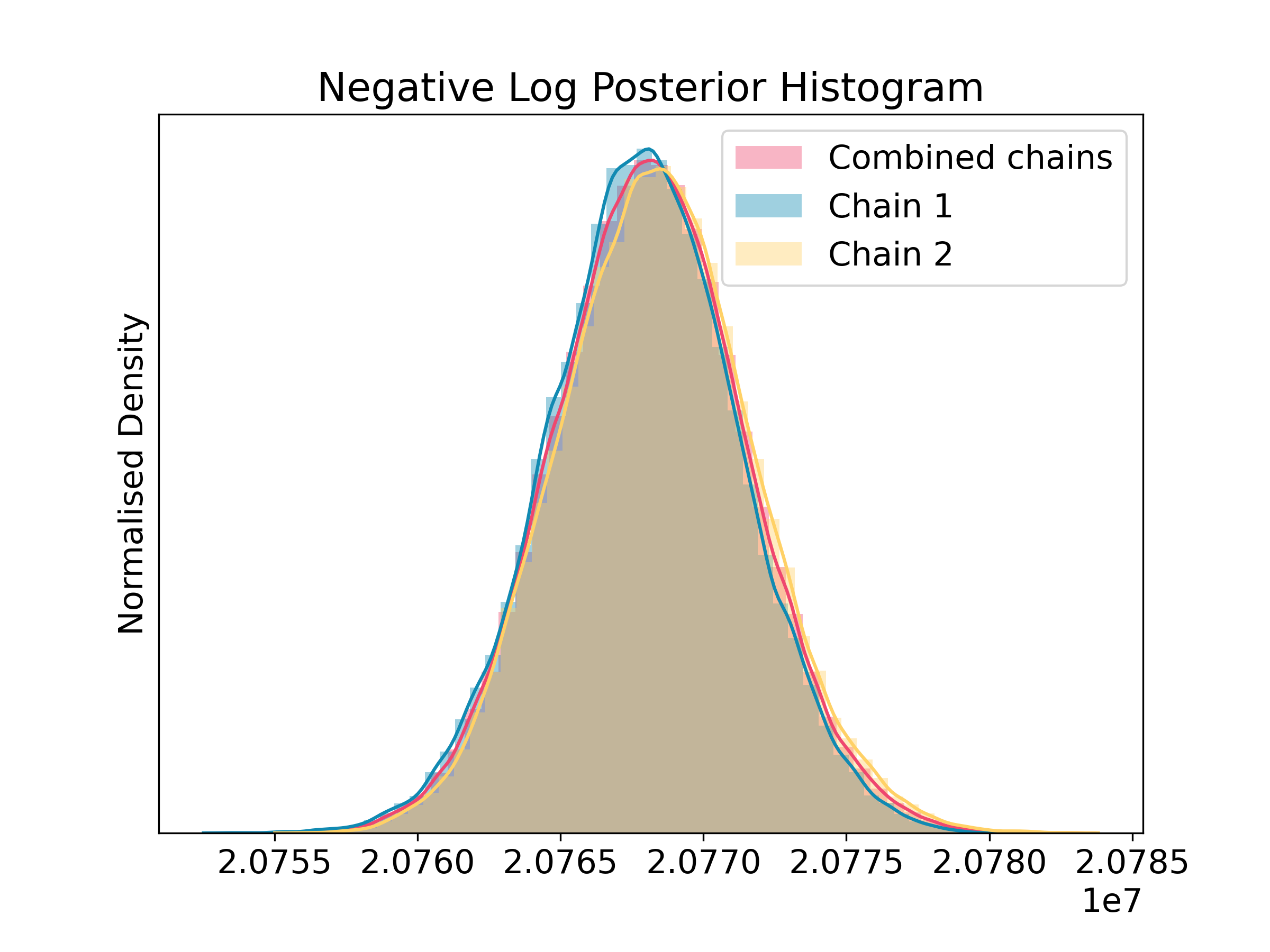} }}%
    \caption{Diagnostics for the log posterior for the two chains (both individually and combined). (a) The log posterior trace for the combined (i.e., concatenated) chain (blue region is chain 1, orange region is chain 2). (b) Combined (pink) and individual chain (blue and orange) histograms of the log posterior values. }%
    \label{Fig:LogPost}%
\end{figure}

Fig.~\ref{Fig:LogPost} shows the negative log posterior trace and its histogram for the combined chains. The trace (upper plot) shows no obvious correlations, and both chains show visually similar behaviour. This first test demonstrates that both chains are consistently probing the same underlying distribution. The histogram of posterior values (bottom plot) also shows a well-behaved distribution, suggesting that burn-in was properly removed for both chains and that both chains burned-in to the same region of the underlying posterior.

\begin{figure}%
    \centering
    \subfloat[\centering Chain 1]{{\includegraphics[width=.45\textwidth]{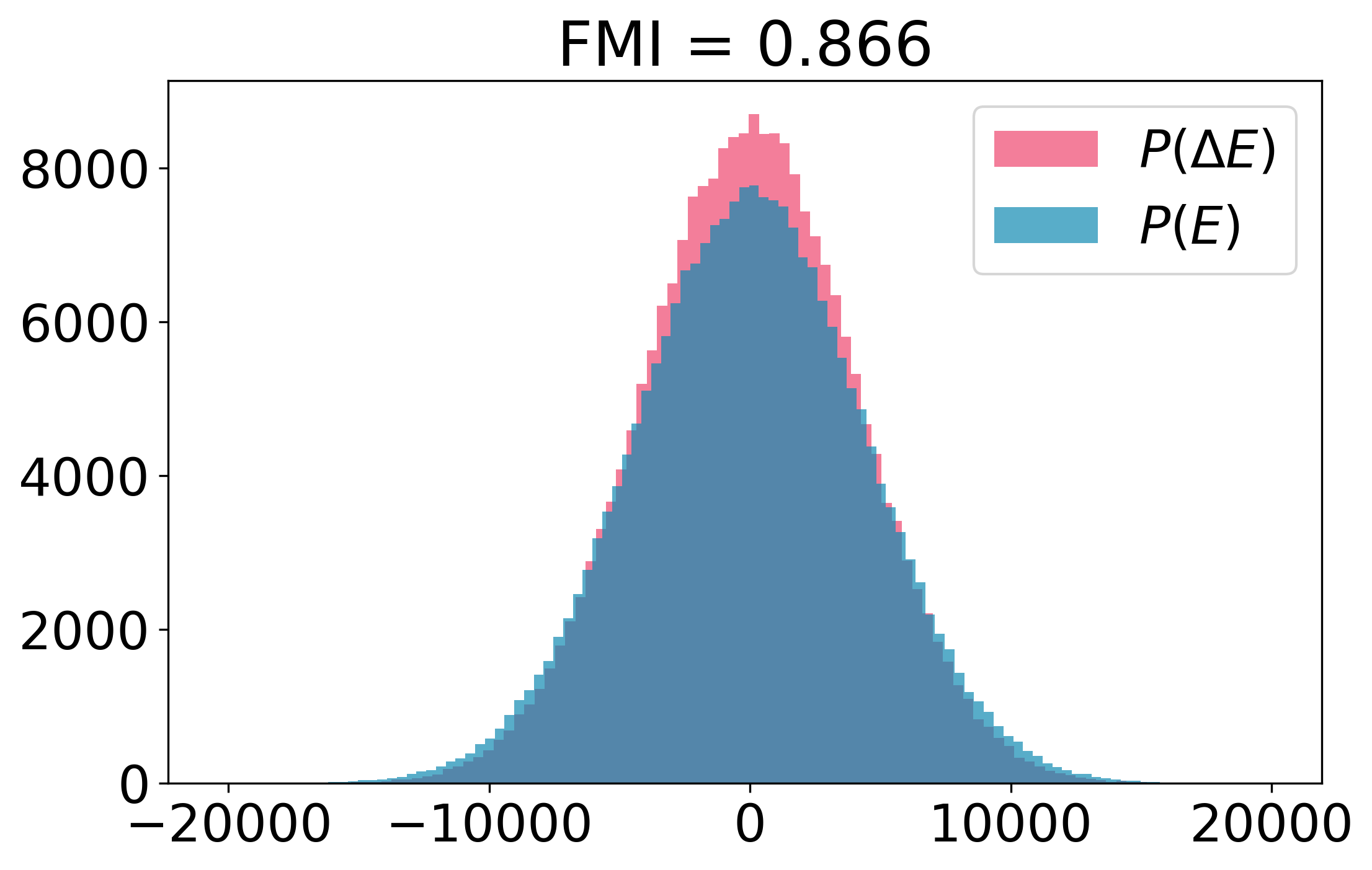} }}%
    \qquad
    \subfloat[\centering Chain 2]{{\includegraphics[width=.44\textwidth]{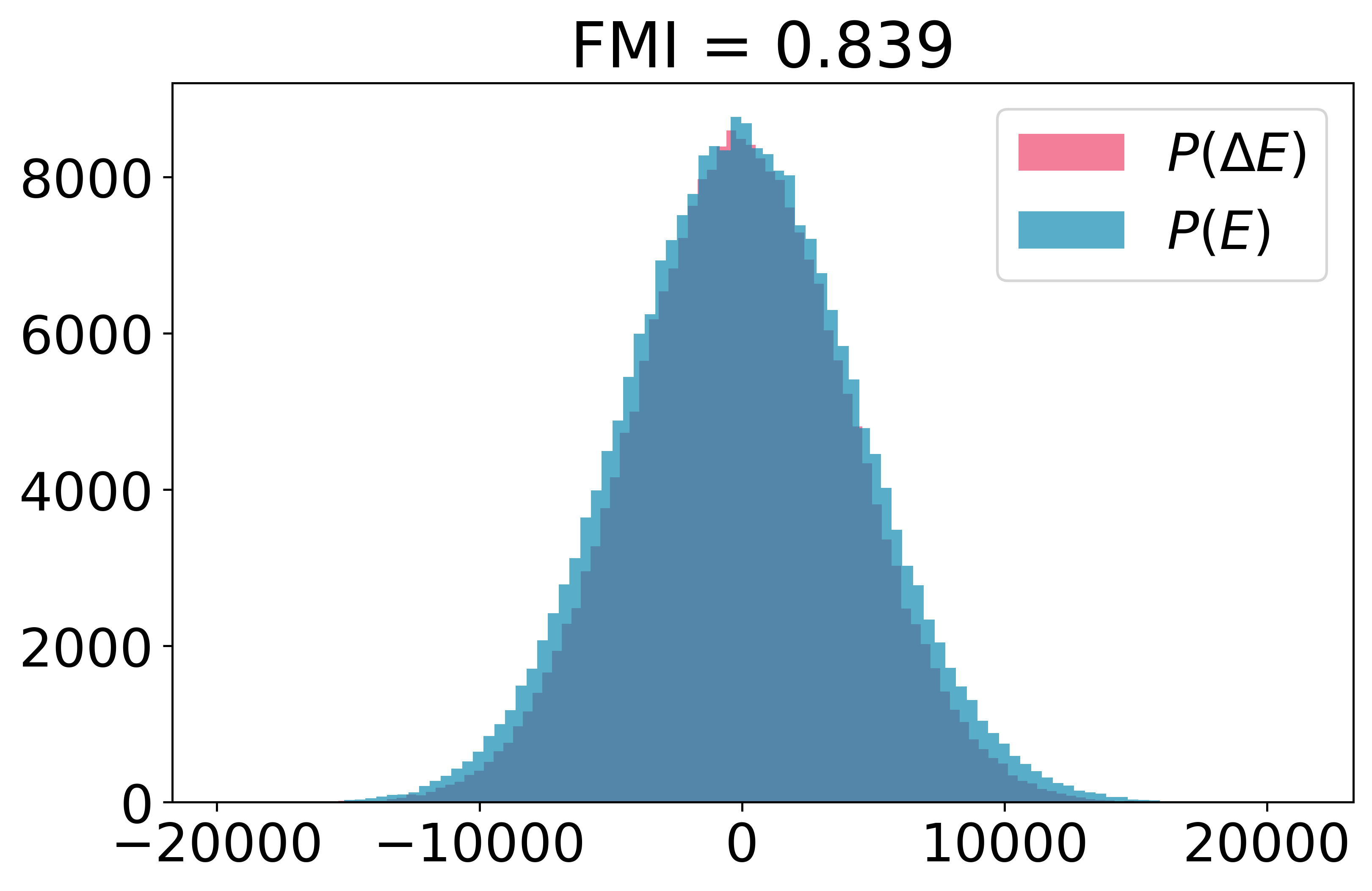} }}%
    \caption{Fraction of Missing Information histograms for the two chains. The two histograms display a similar distribution for both chains.}%
    \label{Fig:FMI}%
\end{figure}

As a second test, we calculate the Fraction of Missing Information (FMI; \citealt{2016-Betancourt-FMI-1}) \footnote{This statistic has also been called the Bayesian Fraction of Missing Information (BFMI); however, the statistic is not specifically Bayesian and hence the shorter name is preferred.}. This diagnostic assesses if the sampler is correctly exploring the full range of total energy submanifolds within phase space. It is defined as \citep{2016-Betancourt-FMI-1,Betancourt}
\begin{equation}
    \mathrm{FMI} =  \frac{ \sum_{n=1}^N (E_n -E_{n-1})^2}{\sum_{n=0}^N (E_n - \bar{E})^2    },
    \label{Eq:FMI}
\end{equation}
where $E_n$ is the total energy of the Hamiltonian system at the \textit{n}th sample in the chain and $\bar{E}$ is the mean energy of the Hamiltonian system for the whole chain. 
In our experience with verifying against other convergence diagnostics (correlation lengths, acceptance ratio, and the Hanson statistic), an FMI $> 0.7$, coupled with similar histograms in Fig.~\ref{Fig:FMI}, indicates that the sampler’s hyperparameters and the coordinate system used are such that the sampler is properly exploring the posterior distribution. At the same time, an FMI $< 0.7$ is associated with poor convergence. Helpfully, the FMI can diagnose bad behaviour with fewer samples than other statistics. For the chains presented in Sec.~\ref{Sec:Results}, we achieve a FMI of 0.87 for chain 1 and 0.84 for chain 2.

The requirement that FMI should be close to unity can be visualised by looking at the histograms of $E= E_n - \bar{E}$ and $\Delta E = E_n -E_{n-1}$ shown in Fig.~\ref{Fig:FMI} for both chains individually. Ideally the histograms of $P(\Delta E)$ and $P(E)$ would be identical, as this would demonstrate that the sampler has correctly explored the full range of available energies within phase space. Even with a very low S/N for the $B$-modes, Fig.~\ref{Fig:FMI} shows that the coordinate transformation discussed in Sec.~\ref{sec:cholesky} allows the HMC sampler to properly explore phase space \citep{2016-Betancourt-FMI-1,Betancourt}.


As a third test we  compute the correlation length of individual chains; this is defined, for each parameter, to be the length at which the auto-correlation of the parameter values in the chain drops below a threshold of $0.1$ for the first time. We find the chains have a median correlation length of $3042$ and $2585$, respectively, which corresponds to having a total of $165$ fully independent samples in the chains. Both these tests demonstrate significant convergence for the individual chains, meaning that the two chains can be combined for more robust constraints.


As a fourth test we examine Hanson's convergence statistic \citep{2001-Hanson-HMC,Taylor},

\begin{equation}
    {\cal H}_i = \frac{
        \frac{1}{3N} \sum_n (x_{n,i} - \bar{x}_i)^3\nabla_{x_i}\nlp{}
    }{
        \frac{1}{N} \sum_{n=1}^N (x_{n,i} - \bar{x}_i)^2
    }\, ,
    \label{Eq:Hanson}
\end{equation}
where $x_{n,i}$ is the $n$th sample of the $i$th parameter. Our $\cal H$ equals that in \cite{2001-Hanson-HMC} and is the reciprocal of that in \cite{Taylor}. For a distribution with exponentially decaying tails, ${\cal H}_i$ is expected to tend to unity for all dimensions $i$, if convergence is reached.

\begin{figure}%
    \centering
    \subfloat[\centering Chain 1]{{\includegraphics[width=.45\textwidth]{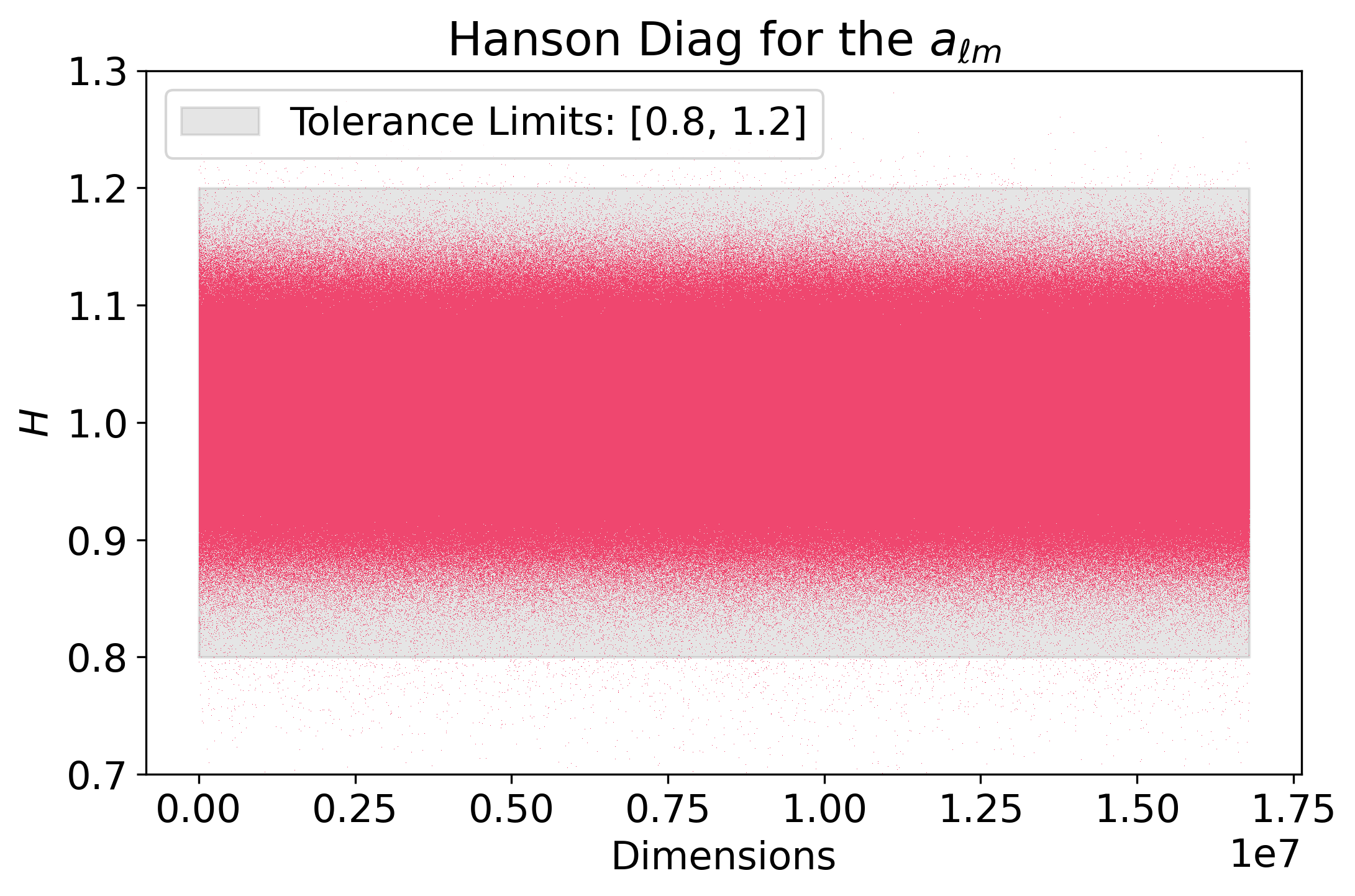} }}%
    \qquad
    \subfloat[\centering Chain 2]{{\includegraphics[width=.45\textwidth]{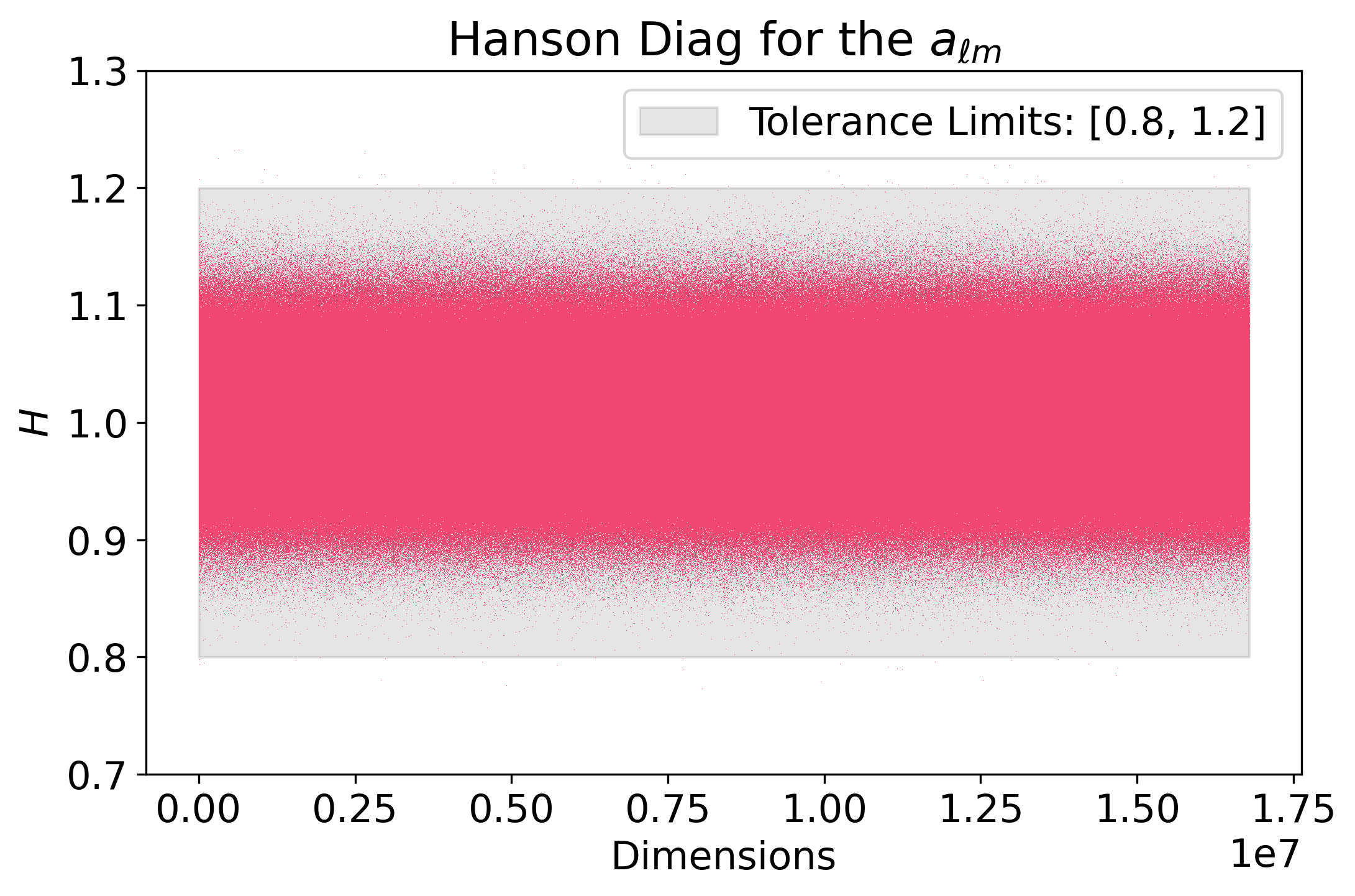} }}%
    \caption{Hanson diagnostics for the spherical harmonic coefficients. The quoted tolerance limits suggest good convergence \citep{Taylor}.} 
    \label{Fig:HansonAlm}%
\end{figure}

\cite{Taylor} states that in practice, ${\cal H}_i$ between 0.8 and 1.2 suggests good convergence in the chains. Fig.~\ref{Fig:HansonAlm} shows this diagnostic applied to the spherical harmonic dimensions of the posterior; this demonstrates good convergence.


\begin{figure}
    \centering 
    \includegraphics[width=0.49\textwidth]{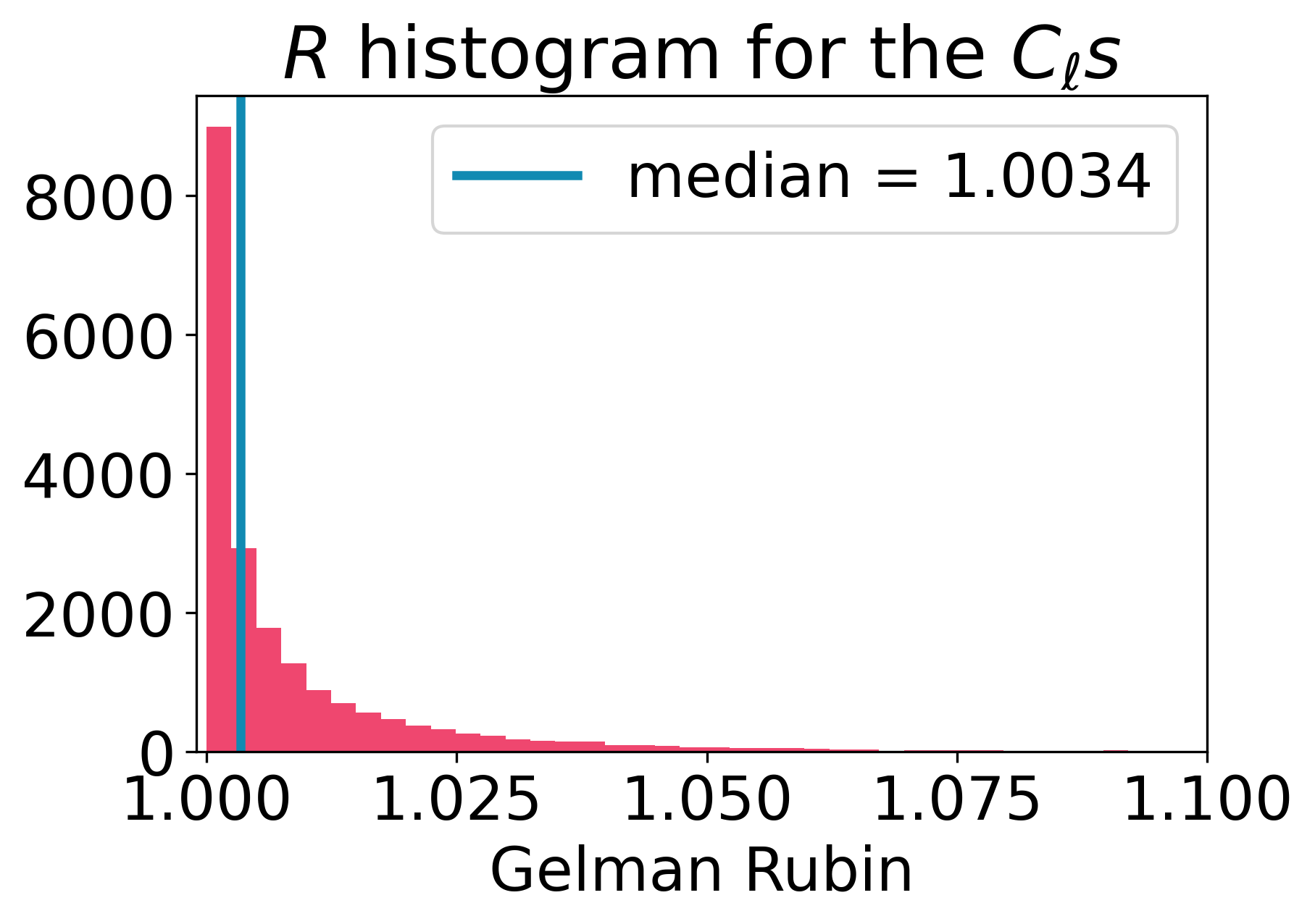}
    \caption{Distribution of Gelman-Rubin $R$ statistics, for the power spectra, for $\ell_{\rm max}=2048$. A very small number of $R$ values are larger than the range shown.}
    \label{Fig:GelmanRubin}
\end{figure}

Finally, since we have two independent chains with different starting points, we can perform a Gelman-Rubin test to further assess convergence. The results of this test are shown in Fig.~\ref{Fig:GelmanRubin} for each of the angular power spectrum dimensions. The median is very close to unity (1.0034) with a small tail to higher values. Although excellent convergence is not achieved for all dimensions, the overall distribution of marginalised angular power spectra demonstrates a satisfactory convergence with $R<1.1$ for the vast majority of parameters and $R<1.4$ when considering all dimensions. Hence, we are confident that the chains have converged to the correct posterior distribution.

\section{Discussion}\label{Sec:Discussion}
This work presents a Bayesian hierarchical model for joint inference of angular power spectra and maps applied to simulated Stage IV tomographic cosmic shear data distributed on the sphere. The method hierarchically models the fields and their covariances, with a harmonic space parameterisation and a Cholesky decomposition of the power spectra. 

This Bayesian approach has the advantage of producing samples of the posterior distribution of the power spectra and map coefficients, thereby avoiding the simplifying assumption of gaussianity that is a feature of summary statistics (and that may or may not be accurate). The posterior distribution has more than 16.8 million parameters, and is thus a challenge for a standard Bayesian approach to joint inference. Such a high-dimensional posterior requires an efficient sampler; we have developed a Hamiltonian Monte Carlo sampler with tuning stages to optimise the step size and the acceptance ratio \citepCore{}. To deal with the low S/N intrinsically found in the $B$-modes (expected to be zero for the standard cosmological model and usual extensions), we performed a coordinate transformation to both the spherical harmonic coefficients, by re-scaling them to their variance, and to the field's covariance in harmonic space --- described in Eqs.~\eqref{Eq:Scale_fata}~and~\eqref{Eq:DiagonalLog_L}.

Although past works have applied similar methods to cosmic shear data \citep{2016-Alsing, 2017-Alsing} and CMB lensing \citep{Millea2021}, this is the first time this methodology has been developed for a complete lensing sky, without the flat-sky approximation. Moving away from the flat-sky approximation is crucial for analysing the next generation of cosmic shear surveys due to the significant increase in sky coverage compared to the previous generations \citep{2017-Lemos-FlatSky, 2020-TutusausEuclid, 2021-Matthewson-FlatSky}. For the method presented here, we can use the increase in sky coverage to our advantage: as more modes are available in the observed data, \almanac{} can robustly infer the missing data under the mask and its angular power spectra.

To test the methodology, we simulated a Euclid-like survey using gaussian simulations to control the simulation's signal, noise, and underlying angular power spectra. The simulations contained two tomographic bins with the expected Euclid mask, galaxy density, and shape noise. We then analysed the cosmic shear simulated data with two chains using different starting points over the multipole range $4 \le \ell \le 2048$. From this analysis, we inferred the angular power spectra of cosmic shear $E$- and $B$-modes from the simulated data. Here, we highlight the highly non-gaussian posterior distributions shown in Fig.~\ref*{Fig:EMode_Hist} for the largest scales and how, even for scales larger than the ones available in data, we can still accurately infer the angular power spectra of the $E$-modes. Although the largest scales in the angular power spectra of cosmic shear carry little cosmological information, these scales allow us to reconstruct the inferred denoised shear maps, as shown in Sect. \ref{SSec:Maps}. We note, however, that such scales are extremely valuable for galaxy clustering studies with primordial non-gaussianities \citep{2022-Andrews-BORG-fnl,2022-Baumann-fnl}. We will study the effects of the non-gaussianities on the angular power spectra in a future work.

Since the method presented here also infers the spherical harmonic coefficients for the underlying signal, we can convert these samples into recovered signal maps for cosmic shear, convergence, and lensing potential. The maps we recover in each sample contain information inferred under areas where the telescope has not collected data. When naively analysing typical maps in the chains, we can recover a significant amount of information, but not sufficient to precisely recover the full signal for small scales (missing due to the mask). Nevertheless, in future works we will explore how the mass maps we recovered, under a gaussian signal assumption, compare to other mainstream mass mapping methods in the literature \citep{1993-Kaiser-MassMapping,2002-Marshall-MassMap,2016-Lanusse-MassMap,2018-Chang-MassMapDES,2020-Price-MassMap,2021-Jeffrey-DES-MASSMAP}.

Differently from (pseudo) angular power spectrum estimators \citep{2019-Alonso-Namaster,2021-Nicola,2004-Chon-PolSpice} or correlation function estimators \citep{TreeCorr,2014-Kilbinger-Athena}, here we are probing the full posterior distribution of full-sky $E$- and $B$-modes $\sfC_{\ell}$. This allows us to inspect for systematic contamination in parity-violating and $B$-modes. For point estimates, the mask can make it difficult to disambiguate between $E$-mode power and $B$-mode power. In some analyses this is described as `$EB$ leakage' for which various solutions are proposed \citep{SEK10,2017-Leistedt,2019-Liu,2021-Gosh-EB-Leak}. With \almanac{}, on the other hand, we have the full posterior and such leakage does not occur. The ambiguity may simply show up as correlation between the $E$ and $B$ mode power spectra (and wider error bars on each power spectrum element individually), and we typically find the marginal distributions of $B$-mode power spectra to have a mode at zero, as we expect.  Only if these posteriors are converted to point estimates would one expect such $EB$ leakage to occur.

In comparison with other BHMs for field-level inference, such as \textsc{Borg} \citep{2013-Jasche-BORG1,2019-Jasche-BORG2,2021-Porqueres-WL-1,2022-Porqueres-2}, \almanac{} has the advantage of being agnostic to the underlying gravitational model, so the map and power spectra samples may be used to test many cosmological models. However, without a specific gravity model included, \almanac{} does not sample the initial conditions of the Universe. Other approaches use a log-normal prior, but do not yet analyse full-sky tomography \citep{Fiedorowicz2022,Boruah2022}.

In terms of computational resources for future Euclid and LSST applications, the method is limited by the spherical harmonic transformations performed using \textsc{libsharp2} \citep{2013-Libsharp,2018-Libsharp2}. For the set-up we presented in this work, we are able to produce 100 samples from the posterior distribution in less than 40 minutes on an {Apple M1 Pro} laptop\footnote{Note, however, that the analysis was not performed on a laptop, but on local cluster computers.}. 
Future investigations of parallelising the posterior calculations over redshift tomographic bins could potentially provide a significant speed-up in the calculations.

Future works will investigate extensions of \almanac{} to non-gaussian likelihoods and to the extraction of cosmological parameters via different approaches, including naively using the marginalised samples and their covariances for inference. We will also explore the use of the \almanac{} to galaxy clustering and to joint analysis of cosmic shear, galaxy clustering, and CMB.

\section*{Acknowledgments}

We thank the support staff of Leiden University's ALICE High Performance Computing infrastructure and UCL's Hypatia cluster (especially Edd Edmondson), Martin Reinecke for assistance with libsharp \citep{2013-Libsharp}, and Malak Olamaie, Sreekumar Balan, Mike Hobson, Joris Bierkens and Justin Alsing for useful discussions. \almanac{} was created by making significant modifications to a software package originally written by Sreekumar Balan and we record our gratitude to him for making this code available to us. The work we present here would not have been possible without the following packages and software: \textsc{NumPy} \citep{harris2020numpy}, \textsc{SciPy} \citep{2020SciPy-NMeth}, \textsc{Matplotlib} \citep{Hunter:2007-matplotlib}, \textsc{HEALPix} \citep{1999-Healpix}, \textsc{libsharp 2} \citep{2018-Libsharp2}, \textsc{GetDist} \citep{2019-Lewis-GetDist}, \textsc{Pandas} \citep{mckinney-proc-scipy-2010}, \textsc{AstroPy} \citep{astropy:2018}, and \textsc{Jupyter Lab} \citep{jupyter}. This work used computing equipment funded by the Research Capital Investment Fund (RCIF) provided by UKRI, and was partially funded by the UCL Cosmoparticle Initiative.

\bibliographystyle{mnras}
\bibliography{Almanac.bib}

\end{document}